\documentclass[journal]{IEEEtran}

\usepackage{cite}
\usepackage[dvips]{graphicx}
\usepackage[cmex10]{amsmath}
\usepackage{color}
\usepackage{bm}
\usepackage{theorem}
\usepackage{amscd}
\usepackage{amssymb}

{\theorembodyfont{\normalfont}
\theoremheaderfont{\normalfont\it}
\newtheorem{thm}{ Theorem}
\newtheorem{dfn}[thm]{ Definition}
\newtheorem{lmm}[thm]{ Lemma}

\newtheorem{crl}[thm]{ Corollary}
\newtheorem{asm}[thm]{ Assumption}
\newtheorem{prp}[thm]{ Proposition}
\newtheorem{cjt}[thm]{ Conjecture}}

{\theorembodyfont{\normalfont}
\theoremheaderfont{\normalfont\it}
\newtheorem{prf}{ Proof:}}

{\theorembodyfont{\normalfont}
\theoremheaderfont{\normalfont\it}
}

{\theorembodyfont{\normalfont}
\theoremheaderfont{\normalfont\it}
}

{\theorembodyfont{\normalfont}
\theoremheaderfont{\normalfont\it}
}

{\theorembodyfont{\normalfont}
\theoremheaderfont{\normalfont\it}
}

{\theorembodyfont{\normalfont}
\theoremheaderfont{\normalfont\it}
}

{\theorembodyfont{\normalfont}
\theoremheaderfont{\normalfont\it}
\newtheorem{rmk}{ Remark.}}

\newcommand{\bra}[1]{\mbox{$\langle#1|$}}

\newcommand{\ket}[1]{\mbox{$|#1\rangle$}}

\newcommand{\inpro}[2]{\mbox{$\left\langle#1|#2\right\rangle$}}

\newcommand{\proj}[1]{\mbox{$\ket{#1}\!\bra{#1}$}}

\newcommand{\norm}[1]{\|#1\|}

\newcommand{\alg}[1]{\begin{align}#1\end{align}}

\newcommand{\nn}{\nonumber}
\newcommand{\ca}[1]{{\mathcal #1}}
\newcommand{\mbb}[1]{{\mathbb #1}}
\newcommand{\mfk}[1]{{\mathfrak #1}}

\newcommand{\bthm}[1]{\begin{thm}\label{thm:#1}}
\newcommand{\ethm}{\end{thm}}

\newcommand{\rThm}[1]{Theorem \ref{thm:#1}}
\newcommand{\blmm}[1]{\begin{lmm}\label{lmm:#1}}
\newcommand{\elmm}{\end{lmm}}
\newcommand{\rLmm}[1]{Lemma \ref{lmm:#1}}

\newcommand{\bdfn}[1]{\begin{dfn}\label{dfn:#1}}
\newcommand{\edfn}{\end{dfn}}

\newcommand{\rDfn}[1]{Definition \ref{dfn:#1}}
\newcommand{\basm}[1]{\begin{asm}\label{asm:#1}}
\newcommand{\easm}{\end{asm}}
\newcommand{\bprp}[1]{\begin{prp}\label{prp:#1}}
\newcommand{\eprp}{\end{prp}}

\newcommand{\rPrp}[1]{Proposition \ref{prp:#1}}
\newcommand{\bcrl}[1]{\begin{crl}\label{crl:#1}}
\newcommand{\ecrl}{\end{crl}}

\newcommand{\bcjt}[1]{\begin{cjt}\label{cjt:#1}}
\newcommand{\ecjt}{\end{cjt}}

\newcommand{\bprf}{\begin{prf}}
\newcommand{\eprf}{\end{prf}}
\newcommand{\brmk}{\begin{rmk}}
\newcommand{\ermk}{\end{rmk}}
\newcommand{\laeq}[1]{\label{eq:#1}}
\newcommand{\req}[1]{(\ref{eq:#1})}

\newcommand{\QED}{\hfill$\blacksquare$}
\newcommand{\lsec}[1]{\label{sec:#1}}

\newcommand{\rSec}[1]{Section \ref{sec:#1}}
\newcommand{\lapp}[1]{\label{app:#1}}
\newcommand{\rapp}[1]{\ref{app:#1}}
\newcommand{\rApp}[1]{Appendix \ref{app:#1}}

\newcommand{\bitem}{\begin{itemize}}
\newcommand{\entem}{\end{itemize}}
\newcommand{\benum}{\begin{enumerate}}
\newcommand{\ennum}{\end{enumerate}}
\newcommand{\bb}{\mathbb}
\newcommand{\otm}{\otimes}

\newcommand{\rFig}[1]{Figure \ref{fig:#1}}

\begin{document}

\title{Communication Cost for Non-Markovianity of Tripartite Quantum States:\\A Resource Theoretic Approach}

\author{Eyuri Wakakuwa
\thanks{E. Wakakuwa is with the Department of Communication Engineering and Informatics, Graduate School of Informatics and Engineering, The University of Electro-Communications, Japan (email: wakakuwa@quest.is.uec.ac.jp). This work is supported by by JSPS KAKENHI, Grant Number 18J01329.}
}

\maketitle

\begin{abstract}
To quantify non-Markovianity of tripartite quantum states from an operational viewpoint, we introduce a class $\Omega^*$ of operations performed by three distant parties. A tripartite quantum state is a free state under $\Omega^*$ if and only if it is a quantum Markov chain. 
We introduce a function of tripartite quantum states that we call the {\it non-Markovianity of formation}, and prove that it is a faithful measure of non-Markovianity, which is continuous and monotonically nonincreasing under a subclass $\Omega$ of $\Omega^*$. We consider a task in which the three parties generate a non-Markov state from scratch by operations in $\Omega$, assisted with quantum communication from the third party to the others, which does not belong to $\Omega$. We prove that the minimum cost of quantum communication required therein is asymptotically equal to the regularized non-Markovianity of formation. 
Based on this result, we provide a direct operational meaning to a measure of bipartite entanglement called the {\it c-squashed entanglement}.
\end{abstract}


\section{Introduction}

The conditional quantum mutual information (CQMI) is defined for a tripartite quantum state and quantifies the amount of correlation between two subsystems that exists when conditioned by the third one. 
CQMI has operational meanings in the context of quantum state redistribution \cite{yard2009optimal,devetak2008exact}, conditional decoupling \cite{berta2018conditional,berta2018deconstruction} and recoverability \cite{fawzi2015quantum}. 
States for which the CQMI is zero are called {\it quantum Markov chains} \cite{hayden04}, and the others are called {\it non-Markov states}.
All non-Markov states can be exploited as a resource for the conditional quantum one-time pad \cite{sharma2017conditional}, which provides another operational meaning to CQMI.
Refs.~\cite{christandl04,brandao11} showed that a bipartite quantum state is entangled if and only if all of its tripartite extensions are non-Markov.
However, the operational understanding of non-Markovianity of quantum states is still limited \cite{gisin2000linking,christandl2004intrinsic,christandl2007unifying}, compared to those of quantum Markov chains (see e.g.~\cite{sutter2018approximate}), entanglement \cite{plenio07,horodecki2009quantum} and that of classical ones \cite{maurer1996towards,maurer1997intrinsic,maurer1999unconditionally,renner2003new,horodecki2005information,banerjee2015secret,chitambar2015classical,chitambar2015distributions}.

The concept of an {\it operational resource theory} (ORT)
has been applied to various notions in quantum information theory, such as coherence, asymmetry, athermality and non-Gaussianity (see \cite{chitambar2018quantum} for a review). 
The approaches based on ORT not only provide an understanding of these notions from an operational viewpoint, but also lead to findings of tasks for which these properties can be exploited as resources.
 In every ORT, states of the system are classified as either {\it free states} or {\it resource states}, and operations therein are classified as either {\it free operations} or {\it non-free operations}. 
 It is required that (i) any free state is generated from scratch by a free operation, and that (ii) any free operation keeps the set of free states invariant.
The main goal of an ORT is to obtain conditions under which a resource state is convertible to another by means of a free operation.

In this paper, we develop an approach in \cite{wakakuwa2017operational} to analyze non-Markovianity of tripartite quantum states from the viewpoint of ORT. We  introduce a class of operations performed by three distant parties, say Alice, Bob and Eve, which we denote by $\Omega^*$. The class $\Omega^*$ consists of public communication among the parties, quantum communication from Alice and Bob to Eve, local operations by each of Alice and Bob, and local {\it reversible} operations by Eve. 
We prove that the set of quantum Markov chains and $\Omega^*$ satisfy the conditions for free states and free operations, namely, Conditions (i) and (ii) presented above. 
 Thereby we provide a groundwork for an ORT of non-Markovianity.

For evaluating non-Markovianity of tripartite quantum states, we introduce a function that we call the {\it non-Markovianity of formation} (nMF). We prove that nMF is a faithful measure of non-Markovianity, which is asymptotically continuous and monotonically nonincreasing under $\Omega$, a subclass of $\Omega^*$ that was introduced in \cite{wakakuwa2017operational}. 
An operational meaning of nMF is investigated in terms of a task that we call {\it non-Markovianity generation}. The task is for the three parties to generate a non-Markov quantum state from scratch by operations in $\Omega$ and quantum communication from Eve to the others, which does not belong to $\Omega$. 
We consider an asymptotic limit of infinitely many copies and vanishingly small error. 
We analyze the {\it non-Markovianity cost}, namely, the minimum cost of quantum communication per copy required for non-Markovianity generation.
We prove that the non-Markovianity cost is equal to  the reguralized nMF. 

A measure of entanglement of a bipartite quantum state called the {\it c-squashed entanglement} \cite{tucci2002entanglement,nagel2003another,yang2009squashed} is obtained from nMF by taking the infimum over all tripartite extensions,
analogously to the {\it squashed entanglement} obtained from CQMI \cite{christandl04}.
Based on the result of non-Markovianity generation, we prove that the regularized c-squashed entanglement is equal to the minimum cost of classical communication required for a task that we call {\it assisted entanglement dilution}. Thereby we provide a direct operational meaning to the c-squashed entanglement.

This paper is organized as follows. 
In \rSec{settings}, we introduce the class $\Omega^*$ of operations by the three parties. We prove that CQMI is monotonically nonincreasing under $\Omega^*$, and that quantum Markov chains are free states under $\Omega^*$.
\rSec{nMF} provides a definition and properties of nMF.  
In \rSec{disdil}, we introduce the task of non-Markovianity generation.
We prove that the non-Markovianity cost of a tripartite state is asymptotically equal to the regularized nMF. 
\rSec{csquashed} analyzes an operational meaning of the c-squashed entanglement.
Conclusions are given in \rSec{discussion}.
Some of the proofs of the main results are provided in appendices.\\

\noindent{\it Notations:} 
A Hilbert space associated with a quantum system $A$ is denoted by ${\mathcal H}^A$, and its dimension is denoted by $d_A$. 
A system composed of two subsystems $A$ and $B$ is denoted by $AB$. 
When $M$  and $N$ are linear operators on ${\mathcal H}^A$ and ${\mathcal H}^B$, respectively, we denote $M\otimes N$ as $M^A\otimes N^B$ for clarity. We abbreviate $|\psi\rangle^A\otimes|\phi\rangle^B$ as $|\psi\rangle^A|\phi\rangle^B$. The identity operator on a Hilbert space is denoted by $I$. We denote $(M^A\otimes I^B)\ket{\psi}^{AB}$ as $M^A\ket{\psi}^{AB}$, and $(M^A\otimes I^B)\rho^{AB}(M^A\otimes I^B)^{\dagger}$ as $M^A\rho^{AB}M^{A\dagger}$. 
The identity operation on a system is denoted by $\rm id$.
When ${\mathcal E}$ is a quantum operation on $A$, we denote $({\mathcal E}\otimes{\rm id}^B)(\rho^{AB})$ as $({\mathcal E}^A\otimes{\rm id}^B)(\rho^{AB})$ or ${\mathcal E}^A(\rho^{AB})$.  For $\rho^{AB}$, $\rho^{A}$ represents ${\rm Tr}_B[\rho^{AB}]$. We denote $|\psi\rangle\!\langle\psi|$ simply as $\psi$. A system composed of $n$ identical systems of $A$ is denoted by $A^n$ or $\bar{A}$, and the corresponding Hilbert space is denoted by $({\mathcal H}^A)^{\otimes n}$ or ${\mathcal H}^{\bar A}$. The Shannon entropy of a probability distribution is denoted as $H(\{p_j\}_j)$, and the von Neumann entropy of a state $\rho^A$ is interchangeably denoted by $S(\rho^A)$ or $S(A)_\rho$.  $\log{x}$ represents the base $2$ logarithm of $x$. For the properties of quantum entropies and mutual informations, see e.g.~\cite{wildetext}.

\section{Operational Framework}\lsec{settings}

In this section, after reviewing the concept of operational resource theory (ORT), we introduce a class $\Omega^*$ of operations performed by three distant parties. 
We prove that the conditional quantum mutual information is monotonically non-increasing under $\Omega^*$, and that a tripartite quantum state is a quantum Markov chain  if and only if it can be generated  from scratch by an operation in $\Omega^*$. 
We also provide examples of operations that do not belong to $\Omega^*$ and can generate non-Markovianity. 
We introduce a subclass $\Omega$ of $\Omega^*$, which will be considered in the remaining sections.

\subsection{General Concepts of Operational Resource Theory}

We briefly review the concept of ORT (see e.g.~\cite{chitambar2018quantum} for the details). 
In an ORT, we consider a system equipped with a certain structure.
For example, one may consider a quantum system composed of several subsystems, a quantum system with a fixed Hamiltonian, or one associated with a symmetry group.
The minimal assumptions that any ORT must satisfy are as follows:
\benum\renewcommand{\labelenumi}{(\alph{enumi})}
\item All operations on the system are classified as either {\it free operations} or {\it non-free operations}.
\item All states of the system are classified as either {\it free states} or {\it resource states}.
\item The set of free states is closed under free operations.
\item Any free state can be generated from scratch by a free operation.
\ennum
The main interest in an ORT is in determining conditions under which a resource state is convertible to another by a free operation.

Depending on which of (a) and (b) is determined prior to the other, 
there are mainly two approaches for constructing an ORT.
The (a)-first approach is logically straightforward, because the classification in (b) is uniquely determined from (a) due to Assumptions (c) and (d).
The (b)-first approach is heuristic in general, because the classification in (a) is not necessarily unique for a given classification in (b).
For example, in entanglement theory in which separable states are regarded as free states, we usually adopt the set of LOCC (local operations and classical communication) for the set of free operations.
However, LOCC is not the only set of operations that satisfies Conditions (c) and (d). It is known that the class of operations called separable operations also satisfies the two conditions (see e.g. \cite{horodecki2009quantum} and the references therein).


Our approach in this paper is the (b)-first one, because we aim at constructing an ORT in which quantum Markov chains are regarded as free states. 
However, for the logical clarity, we adopt the (a)-first approach in presenting the obtained results, in which case the free states are {\it defined as} the states that satisfy Condition (d).

\begin{figure}[t]
\begin{center}
\includegraphics[bb={0 0 562 367}, scale=0.42]{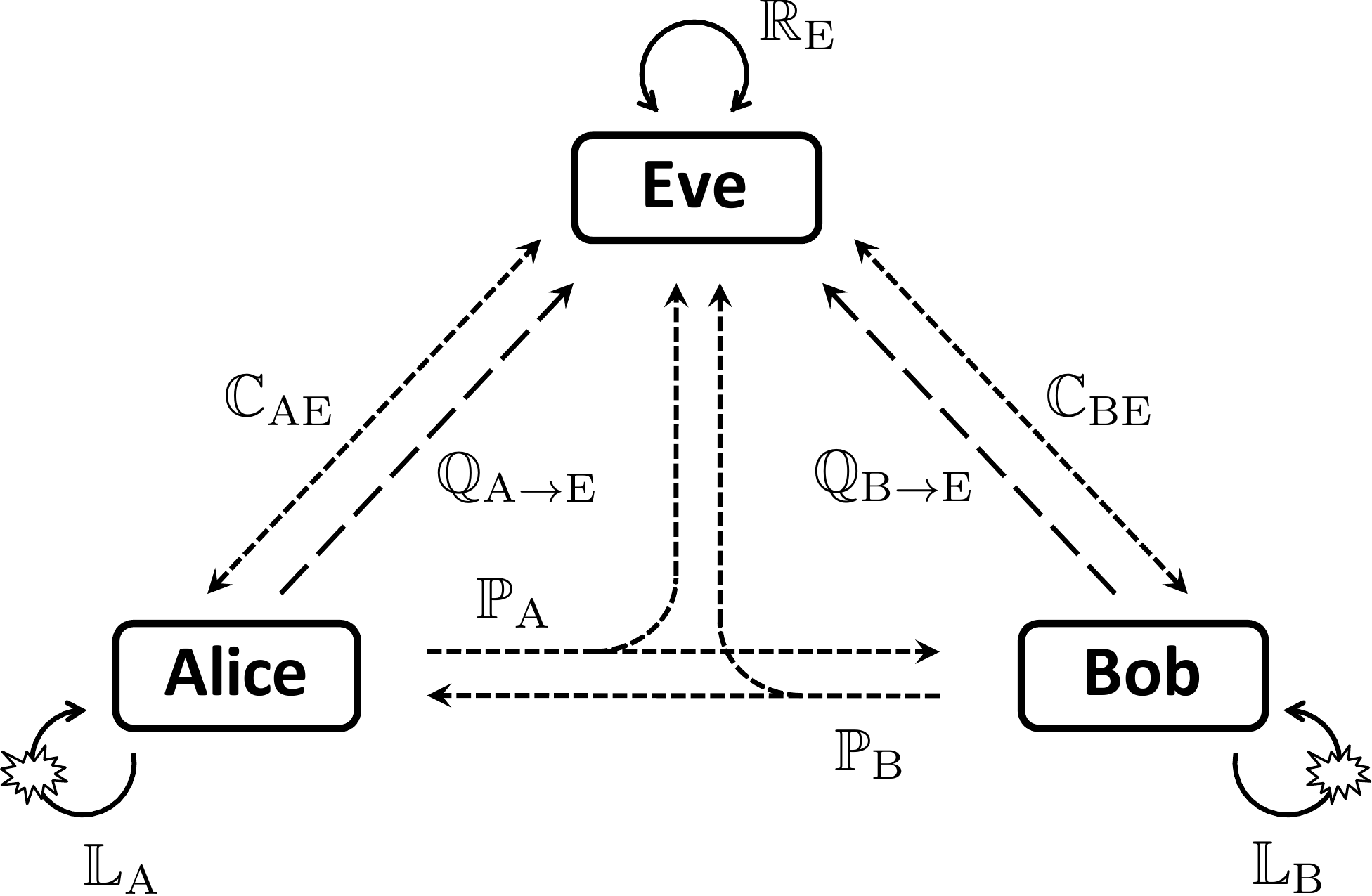}
\end{center}
\caption{The classes of operations that comprises free operations are depicted. Any operation in $\Omega^*$ is represented as a composition of operations in the classes depicted in this figure. We denote classical communication between Alice and Eve simply by $\mbb{C}_{\rm AE}$, and one between Bob and Eve by $\mbb{C}_{\rm BE}$.}
\label{fig:freeoperations}
\end{figure}

\subsection{Free Operations and Convertibility among The States}

Consider three distant parties, say Alice, Bob and Eve. 
We introduce classes of operations performed by the parties.
In order that the obtained ORT has operational significance, we only consider classes of {\it LOCQC} operations, i.e.,  those composed of local operations by each parties, noiseless classical communication and noiseless quantum communication among the parties.
The classes are as follows:
\alg{
{\bb L}_{\rm A}&\text{: local operations by Alice}\nn\\
{\bb L}_{\rm B}&\text{: local operations by Bob}\nn\\
{\bb R}_{\rm E}&\text{: local reversible operations by Eve}\nn\\
{\bb Q}_{\rm A\rightarrow E}&\text{: quantum communication from Alice to Eve}\nn\\
{\bb Q}_{\rm B\rightarrow E}&\text{: quantum communication from Bob to Eve}\nn\\
{\bb P}_{\rm A}&\text{: broadcasting of classical messages by Alice}\nn\\
{\bb P}_{\rm B}&\text{: broadcasting of classical messages by Bob}\nn\\
{\bb C}_{\rm A\rightarrow E}&\text{: classical communication from Alice to Eve}\nn\\
{\bb C}_{\rm E\rightarrow A}&\text{: classical communication from Eve to Alice}\nn\\
{\bb C}_{\rm B\rightarrow E}&\text{: classical communication from Bob to Eve}\nn\\
{\bb C}_{\rm E\rightarrow B}&\text{: classical communication from Eve to Bob}\nn
}
Each operation in ${\bb L}_{\rm A}$, ${\bb L}_{\rm B}$ and ${\bb R}_{\rm E}$ is represented by a linear completely-positive trace-preserving map. The input and output systems of those maps are, in general, composed of classical and quantum subsystems.
An operation $\ca{V}:E\rightarrow \hat{E}$ is said to be {\it reversible} if there exists an operation ${\ca V}^*:\hat{E}\rightarrow E$ such that ${\ca V}^*\circ{\ca V}$ is the identity operation on system $E$, i.e., ${\ca V}^*\circ{\ca V}={\rm id}^E$. 
Examples are unitary operations on $E$ and addition of an ancillary system in a fixed state.
We require that Eve cannot refuse to receive classical messages sent by Alice in ${\mathbb P}_A$ and one by Bob in ${\mathbb P}_B$. This implies that {\it secret} communication channels between Alice and Bob are not available. 
We denote by $\Omega^*$ the set of operations that can be represented as a composition of operations in the above classes (see \rFig{freeoperations}).

In the rest of this paper, we analyze convertibility of tripartite quantum states shared among Alice, Bob and Eve, under $\Omega^*$ and a subclass thereof. The set of all tripartite quantum states is denoted by $\mfk{S}_{\rm all}$.
Due to the condition of reversibility of Eve's operations, it is too restrictive to define convertibility of a state $\rho_1$ to $\rho_2$ by the existence of an operation in $\Omega^*$ that maps $\rho_1$ to $\rho_2$. Thus, we define the convertibility of states under $\Omega^*$ by taking the degree of freedom of reversible operations by Eve into account.
A  rigorous definition is as follows:
\bdfn{exactconvertibility}
A state $\rho_1$ is {\it convertible to $\rho_2$ under $\Omega^*$} if there exist operations $\ca{F}\in\Omega^*$ and ${\ca V}\in\mbb{R}_{\rm E}$ such that
$
\ca{F}(\rho_1^{ABE})={\ca V}(\rho_2^{ABE})
$.
\edfn

\subsection{Monotonicity of The Conditional Quantum Mutual Information}
\lsec{monotCQMI}

For a tripartite quantum state $\rho$ on system $ABE$, the {\it conditional quantum mutual information (CQMI)}  is defined by
\alg{
&I(A:B|E)_\rho\nn\\
&\quad\quad:=S(AE)_\rho+S(BE)_\rho-S(ABE)_\rho-S(E)_\rho.\nn
}
Here, $S$ is the von Neumann entropy of the reduced state of $\rho$ on each subsystem, e.g., 
\alg{
S(AE)_\rho=S(\rho^{AE})=-{\rm Tr}[\rho^{AE}\log{\rho^{AE}}].
}
For simplicity, we denote $\frac{1}{2}I(A:B|E)_\rho$ by $M_I(\rho)$. 
The strong subadditivity of the von Neumann entropy implies that CQMI is nonnegative \cite{lieb1973proof}.
The following lemma states that CQMI is monotonically nonincreasing under $\Omega^*$. 

\blmm{monotCQMI}
For any $\rho\in{\mfk S}_{\rm all}$ and $\ca{F}\in\Omega^*$, it holds that $M_I(\rho)\geq M_I(\ca{F}(\rho))$. 
\elmm

\bprf

It suffices to prove that $M_I$ is monotonically nonincreasing under any class of operations that comprises $\Omega^*$. 
Monotonicity of functions under $\mathbb{C}_{\rm A\rightarrow E}$ immediately follows from one under ${\bb Q}_{\rm A\rightarrow E}$, because communicating a classical message is equivalent to sending a quantum system, under the condition that the state is diagonal with respect to a given basis. 
Monotonicity under ${\bb P}_{\rm A}$ follows from one under $\mathbb{C}_{\rm A\rightarrow E}$ and $\mathbb{C}_{\rm E\rightarrow B}$.
This is because broadcasting of a classical message by Alice is equivalent to Alice's communicating a classical message to Eve, followed by Eve's transferring it to Bob while keeping the copy thereof. 
Monotonicity under ${\bb L}_{\rm A}$, ${\bb P}_{\rm A}$, ${\bb C}_{\rm A\rightarrow E}$, ${\bb C}_{\rm E\rightarrow A}$ and ${\bb Q}_{\rm A\rightarrow E}$ are equivalent to those under ${\bb L}_{\rm B}$, ${\bb P}_{\rm B}$, ${\bb C}_{\rm B\rightarrow E}$, ${\bb C}_{\rm E\rightarrow B}$ and ${\bb Q}_{\rm B\rightarrow E}$, respectively, due to symmetry of CQMI in $A$ and $B$.  
Therefore, we only need to prove that $M_I$ is monotonically nonincreasing under ${\bb L}_{\rm A}$, ${\mathbb R}_{\rm E}$, ${\bb C}_{\rm E\rightarrow B}$ and ${\bb Q}_{\rm A\rightarrow E}$.

\benum

\item {\it Monotonicity under ${\mathbb L}_A$}:
Due to the data processing inequality for CQMI \cite{christandl04}, 
for any $\rho\in{\mfk S}_{\rm all}$ and $\ca{F}\in{\mathbb L}_A$, it holds that
$
I(A:B|E)_{\rho}\geq I(A:B|E)_{\ca{F}(\rho)}
$.

\item {\it Monotonicity under ${\mathbb R}_{\rm E}$}:
Since the quantum mutual information is invariant under local reversible operations, 
we have 
$
I(A:BE)_\rho =I(A:BE)_{{\ca V}(\rho)}\nn
$
and
$
I(A:E)_\rho =I(A:E)_{{\ca V}(\rho)}.\nn
$
Thus, by the chain rule of CQMI, we have
$
I(A:B|E)_\rho=I(A:BE)_\rho-I(A:E)_\rho
=I(A:BE)_{{\ca V}(\rho)}-I(A:E)_{{\ca V}(\rho)}
=I(A:B|E)_{{\ca V}(\rho)}
$.

\item{\it Monotonicity under ${\mathbb C}_{\rm E\rightarrow B}$}: 
The states before and after classical communication from Eve to Bob are represented by density operators
\alg{
\rho_i=\sum_mr_m\proj{m}^{M_{\!E}}\otimes\rho_m^{ABE}\laeq{rhohati2}
} 
and
\alg{
\rho_f=\sum_mr_m\proj{m}^{M_{\!B}}\otimes\proj{m}^{M_{\!E}}\otimes\rho_m^{ABE},\laeq{rhohatf3}
} 
respectively. Here, $\{r_m\}_m$ is a probability distribution, $\{\rho_m\}_m$ is a set of quantum states on $ABE$, $\{\ket{m}\}_m$ is a set of orthonormal pure states, and $M_B$ and $M_E$ are ``classical'' systems in which the messages are stored. 
It follows that
$
2M_I(\rho_i)=
I(A:B|EM_E)_{\rho_i}
=
\sum_mr_mI(A:B|E)_{\rho_m}
=
I(A:BM_B|M_EE)_{\rho_f}
=2M_I(\rho_f)
$.

\item{\it Monotonicity under ${\mathbb Q}_{\rm A\rightarrow E}$}: 
Let $Q$ be a quantum system transmitted from Alice to Eve.
By the chain rule of CQMI, we have
$
I(QA:B|E)_{\rho}
=I(Q:B|E)_{\rho}+I(A:B|EQ)_{\rho}
\geq I(A:B|EQ)_{\rho}
$.
\QED

\ennum

\eprf

\subsection{Quantum Markov Chains are Free States}\lsec{freestate}

Tripartite quantum states for which CQMI is zero are called {\it quantum Markov chains} \cite{hayden04}. For a tripartite system composed of systems $A$, $B$ and $E$, the condition is represented as
\begin{eqnarray}
I(A:B|E)_{\xi}=0.\laeq{markovianity}
\end{eqnarray}
We denote the set of quantum Markov chains satisfying \req{markovianity} by ${\mfk S}_{\rm Markov}$. Ref.~\cite{hayden04} proved that Equality (\ref{eq:markovianity}) is equivalent to the condition that there exists a linear isometry $\Gamma$ from $E$ to $E_0E_LE_R$ such that
\begin{eqnarray}
\Gamma\xi^{ABE}\Gamma^\dagger=\sum_{j\in{\ca J}}p_j\proj{j}^{E_0}\otm\varsigma_j^{AE_L}\otimes\tau_j^{BE_R}.\laeq{decomposability}
\end{eqnarray} 
Here, $E_0$, $E_L$ and $E_R$ are finite-dimensional quantum systems,  $\{p_j\}_{j\in{\ca J}}$ is a probability distribution, $\{|j\rangle\}_{j\in{\ca J}}$ is an orthonormal basis of $E_0$, and $\varsigma_j$ and $\tau_j$ are quantum states on systems $AE_L$ and $BE_R$, respectively, for each $j$. 
The following proposition implies that a state is a free state under $\Omega^*$ if and only if it is a quantum Markov chain. Thus $\Omega^*$ provides a groundwork for an ORT of non-Markovianity.

\bprp{freestate}
A state $\sigma^{ABE}$ can be generated from scratch by an operation in $\Omega^*$ if and only if it is a quantum Markov chain. 
That is, the ``dummy state'' $\sigma_0:=|0\rangle\!\langle0|^A\otm|0\rangle\!\langle0|^B\otm|0\rangle\!\langle0|^E$ is convertible to $\sigma^{ABE}$ under $\Omega^*$ if and only if $\sigma\in{\mfk S}_{\rm Markov}$.
\eprp

\bprf
To prove the ``if'' part, consider the following procedure: 
\begin{enumerate}
\item
Alice generates a random variable $J$ which takes values in $\ca J$ according to a probability distribution $\{p_j\}_{j\in\ca{J}}$.
\item
Alice broadcasts $J$ to Bob and Eve.
\item
Eve records $J$ on her register $E_0$.
\item
Alice locally prepares a state $\varsigma_j^{AE_L}$ and sends $E_L$ to Eve.
\item
Bob locally prepares a state $\tau_j^{BE_R}$ and sends $E_R$ to Eve.
\item
Alice and Bob discards $J$. 
\end{enumerate}
It is straightforward to verify that any state in the form of \req{decomposability} can be generated by this protocol. 
Noting that $\Gamma$ is a reversible operation, this completes the proof of the ``if'' part.

To prove the ``only if'' part, suppose that $\sigma\in{\mfk S}_{\rm all}$ is a free state. By definition, for the dummy state $\sigma_0\in{\mfk S}_{\rm Markov}$, there exists an operation $\ca{F}\in\Omega^*$ and a reversible operation $\ca{V}\in\mbb{R}_E$ such that $\ca{F}(\sigma_0)=\ca{V}(\sigma)$.  From the monotonicity of CQMI under $\Omega^*$, it follows that $0=M_I(\sigma_0)\geq M_I(\sigma)$, which yields $M_I(\sigma)=0$ and thus $\sigma\in{\mfk S}_{\rm Markov}$.\QED
\eprf

\subsection{Operations That Can Generate Non-Markovianity}
\lsec{maximality}

Among the classes of operations within LOCQC (local operations and classical and quantum noiseless communication),
the following classes are not included in $\Omega^*$:
\alg{
{\bb L}_{\rm E}\backslash{\bb R}_{\rm E}&\text{: local irreversible operations by Eve}\nn\\
{\bb Q}_{\rm E\rightarrow A}&\text{: quantum communication from Eve to Alice}\nn\\
{\bb Q}_{\rm E\rightarrow B}&\text{: quantum communication from Eve to Bob}\nn\\
{\bb S}_{\rm AB}&\text{: secret communication between Alice and Bob}\nn\\
{\bb Q}_{\rm AB}&\text{: quantum communication between Alice and Bob}\nn
}
The following examples show that operations in the above classes can generate non-Markov state from scratch.

\bitem
\item ${\bb L}_{\rm E}\backslash{\bb R}_{\rm E}$:
Suppose that the three parties initially share a quantum state
\alg{
\quad
\rho:=\frac{1}{2}(\proj{000}^{ABE}+\proj{111}^{ABE})\in\mfk{S}_{\rm Markov}.
\!\!\!
\nn
}
By Eve randomly performing $I$ or $\sigma_x$ with probability $1/2$ on her qubit, the above state is transformed to
\alg{
\rho':=\frac{1}{2}(\proj{00}^{AB}+\proj{11}^{AB})\otm\frac{1}{2}I^E,
} 
which is not a quantum Markov chain because $M_I(\rho')=\frac{1}{2}$. 

\item ${\bb Q}_{\rm E\rightarrow A}$:
Suppose that the initial state is
\alg{
\quad
\rho:=\proj{0}^{A'}\otm\proj{\Phi_2}^{BE'}\otm\proj{0}^{E}\in\mfk{S}_{\rm Markov},\!\!\!\nn
}
where $\ket{\Phi_2}:=(\ket{00}+\ket{11})/\sqrt{2}$ is a Bell state. Eve sends $E'$ to Alice, and then Alice discards $A'$. The obtained state is
\alg{
\rho':=\proj{\Phi_2}^{AB}\otm\proj{0}^{E},
\laeq{maxent}
} 
which is not a quantum Markov chain because $M_I(\rho')=1$.
\item ${\bb Q}_{\rm E\rightarrow B}$:
By exchanging the roles of Alice and Bob in the above example, it follows that quantum communication from Eve to Bob can generate a non-Markov state.

\item ${\bb S}_{\rm AB}$:
Suppose that the three parties initially share a state $\rho:=\proj{000}^{ABE}\in\mfk{S}_{\rm Markov}$. Alice flips a fair coin and sends the result through a secret communication channel to Bob. Depending on the result, Alice and Bob performs $I$ or $\sigma_x$ to their qubits, after which the state is
\alg{
\rho':=\frac{1}{2}(\proj{00}^{AB}+\proj{11}^{AB})\otm\proj{0}^E.
} 
This is not a quantum Markov chain because $M_I(\rho')=\frac{1}{2}$.

\item ${\bb Q}_{\rm AB}$:
It is straightforward to verify that, by quantum communication from Alice to Bob, the three parties can obtain a state in the form of \req{maxent} from scratch.

\entem

\subsection{A Subclass $\Omega$ of $\Omega^*$}
\lsec{NFO}

Let $\Omega$ be the class of operations composed of operations in all classes comprising $\Omega^*$  but ${\bb C}_{\rm E\rightarrow A}$ and ${\bb C}_{\rm E\rightarrow B}$. 
It immediately follows that $\Omega\subset \Omega^*$.
In the rest of this paper, we consider convertibility of tripartite quantum states under operations in $\Omega$ assisted by non-free operations. 
Ideally, it would be desirable to adopt $\Omega^*$ as a class of free operations, instead of $\Omega$. However, at this point, we have not succeeded in proving monotonicity of a function (the non-Markovianity of formation) under $\Omega^*$, which plays a central role in analyzing non-Markovianity generation protocols (see \rSec{nMF} for the detail). Thus we adopt $\Omega$ for free operations in the rest of this paper. It should be noted that a tripartite quantum state is a quantum Markov chain if and only if it is a free state under $\Omega$ (see the proof of \rPrp{freestate} in \rSec{freestate}). The structure of an ORT under $\Omega$ has been analyzed in \cite{wakakuwa2017operational}.

\section{The Non-Markovianity of Formation}
\lsec{nMF}

In this section, we introduce a function of tripartite quantum states that we call the {\it non-Markovianity of formation} (nMF).
We prove its properties such as faithfulness, continuity and monotonicity under $\Omega$. 
An operational meaning of the (regularized) nMF will be provided in \rSec{disdil} in the context of a non-Markovianity generation protocol.

\subsection{Definition}

Consider a tripartite quantum state $\rho$ on system $ABE$.
Suppose that $\rho$ is decomposed in the form of $\rho=\sum_kp_k\rho_k$, where $\{p_k,\rho_k\}_k$ is an ensemble of quantum states on $ABE$. For each $k$, define
\alg{
\!\!
 \lambda(\rho_k):=\!
\inf_{A'B'}\inf_{\tilde{\rho}_k}[S(AA')_{\tilde{\rho}_k}\!+\!S(BB')_{\tilde{\rho}_k}\!-\!S(A'B')_{\tilde{\rho}_k}]. \!\!
\laeq{dfnhatI}
}
The infimum is taken over all finite dimensional quantum systems $A'$, $B'$ and all quantum states $\tilde{\rho}_k$ on $AA'BB'E$ such that ${\rm Tr}_{A'B'}[\tilde{\rho}_k]=\rho_k$, similarly to the definition of the conditional entanglement of mutual information \cite{yang2008additive}. 
Using this function, we define $\Lambda$ by 
\alg{
 \Lambda(\rho):=\inf_{\mathbb{K}}\inf_{\{p_k,\rho_k\}_{k\in\mathbb{K}}}\left[\sum_{k\in\mathbb{K}}p_k \lambda (\rho_k)\right],
}
where the infimum is taken over all finite sets $\mathbb{K}$ and ensembles $\{p_k,\rho_k\}_{k\in\mathbb{K}}$ such that $\rho=\sum_{k\in\mathbb{K}}p_k\rho_k$. 
The {\it non-Markovianity of formation} (nMF) is defined as
\alg{
M_F(\rho):=
\frac{1}{2}[S(AB|E)_\rho+ \Lambda(\rho)],
\laeq{dfnIf}
}
in which the first term is the conditional entropy defined by $S(AB|E)_\rho:=S(ABE)_\rho-S(E)_\rho$.

For later convenience, we present an alternative (but equivalent) expression for nMF. Let $\{p_k,\ket{\phi_k}\}_{k}$ be an arbitrary ensemble of pure states on $AA'BB'EE'$ such that
\alg{
\rho^{ABE}=\sum_{k}p_k{\rm Tr}_{A'B'E'}[\proj{\phi_k}].
\laeq{decpurho}
} 
Consider a state in the form of
\alg{
\!\varrho^{AA'BB'EE'K}
=
\sum_{k}p_k\proj{\phi_k}^{AA'BB'EE'}
\!\otm\proj{k}^K,
\!
\laeq{dfnvarrhoK}
}
where $K$ is a quantum system with an associated orthonormal basis $\{\ket{k}\}_{k}$.
As we prove in \rApp{prfnMF2}, nMF is represented as
\alg{
&M_F(\rho)=\nn\\
&
\frac{1}{2}
\inf_{A',B',E',\mathbb{K}}\inf_{\{p_k,|\phi_k\rangle\}}
\left[I(AA'\!:\!BB'|K)_\varrho+I(AB\!:\!E'K|E)_\varrho\right]\!,\laeq{nMF2}
}
where the infimum is taken over all finite dimensional quantum systems $A'$, $B'$, $E'$, finite sets $\mathbb{K}$ and ensembles $\{p_k,|\phi_k\rangle\}_{k\in\mathbb{K}}$ that satisfy \req{decpurho}. Due to the nonnegativity of the (conditional) quantum mutual information, it immediately follows that $M_F$ is nonnegative.
For clarity, we interchangeably denote nMF by $M_F(\rho)$ or $M_F(A:B|E)_\rho$. By definition, nMF is symmetric in $A$ and $B$, namely, it holds that
\alg{
M_F(A:B|E)_\rho=M_F(B:A|E)_\rho.
\laeq{symnMF}
}

\subsection{Properties}

The following lemma states that nMF is a measure of non-Markovianity of quantum states, which has similar properties as the conditional quantum mutual information. Proofs are provided in \rApp{propnMF} and \rapp{asycomnMF}.

\blmm{propIf}
The non-Markovianity of formation satisfies the following properties:
\bitem
\setlength{\leftskip}{2mm}
\item[(P1)]{\it lower bound}: $M_F(\rho)\geq M_I(\rho)$ for any $\rho\in{\mfk S}_{\rm all}$.
\item[(P2)]{\it upper bound}: $M_F(\rho)\leq \min\{S(A)_\rho,S(B)_\rho\}$ for any $\rho\in{\mfk S}_{\rm all}$.
\item[(P3)]{\it faithfulness}: $M_F(\rho)=0$ if and only if $\rho\in{\mfk S}_{\rm Markov}$.
\item[(P4)]{\it pure states}: $M_F(\psi)=\frac{1}{2}I(A:B)_\psi$ for any pure state $\psi$ on $ABE$.
\item[(P5)]{\it subadditivity}: $ M_F(\rho\otimes\sigma)\leq M_F(\rho)+M_F(\sigma)$ for any $\rho,\sigma\in{\mfk S}_{\rm all}$.
 \item[(P6)]{\it weak chain rule}: For any state on system $ABCE$, 
 it holds that 
 \alg{
M_F(AC:B|E)_\rho
\geq
 M_F(A:B|EC)_{\rho}.
 \nn
 }
\item[(P7)]{\it conditional convexity}: For any ensemble $\{r_m,\rho_m\}_m$ of states on $ABE$, and for the state
\alg{
\rho=\sum_mr_m\rho_m^{ABE}\!\otm\!\proj{m}^{M}\!, 
\;
\inpro{m}{m'}\!=\!\delta_{m,m'},
\!\!\!\!\!\!\!
\laeq{decCQst}
}
 it holds that 
 \alg{
M_F(A:B|EM)_\rho\leq\sum_mr_mM_F(A:B|E)_{\rho_m}.
\!\!\!\!\!\!
 \laeq{convMF}
 }
\item[(P8)]{\it invariance under reversible operations}: For any $\rho\in{\mfk S}_{\rm all}$ and $\ca{V}\in\mathbb{R}_E$, it holds that $M_F({\ca V}(\rho))=M_F(\rho)$.
\item[(P9)]{\it average monotonicity}: For any state $\rho\in{\mfk S}_{\rm all}$ and any measurement on $A$, 
 it holds that 
 \alg{
\quad\quad\quad M_F(A:B|E)_\rho\geq\sum_m\nu_m M_F(A:B|E)_{\rho_m},
 \nn
 }
 where $\nu_m$ is the probability of obtaining the outcome $m$, and $\rho_m$ is the state after the measurement corresponding to the outcome $m$.
\item[(P10)]{\it $\Omega$-monotonicity}: For any $\rho\in{\mfk S}_{\rm all}$ and $\ca{F}\in\Omega$, it holds that $M_F(\rho)\geq M_F(\ca{F}(\rho))$.
\item[(P11)]{\it asymptotic continuity}: For any $\rho,\sigma\in{\mfk S}_{\rm all}$ satisfying $\frac{1}{2}\|\rho-\sigma\|_1\leq\epsilon\leq1$, it holds that $|M_F(\rho)-M_F(\sigma)|\leq 4\sqrt{\epsilon}\log{d_Ad_B}+h'(\epsilon)$, where $h'$ is a function that satisfies $\lim_{\epsilon\rightarrow0}h'(\epsilon)=0$ and is independent of dimensions of the systems.
\entem
\elmm
Due to the subadditivity (P5), the following limit exists:
\alg{
M_F^\infty(\rho):=\lim_{n\rightarrow\infty}\frac{1}{n}M_F(\rho^{\otimes n}).
}
We refer to this function as the {\it regularized non-Markovianity of formation}. It should be noted that $M_I$ also satisfies Properties (P1)-(P11), saturating inequalities in (P1), (P5) and (P7).

In \rSec{disdil}, we will consider a task in which a tripartite quantum state is generated from scratch by operations in $\Omega$ and quantum communication from Eve to Alice and Bob.
We analyze the minimum cost of quantum communication required for the task.
The amount of quantum communication from Eve to Alice and Bob is quantified simply by the number of qubits transmitted. That is, for ${\ca G}\in{\mbb Q}_{\rm E}:={\mbb Q}_{\rm E\rightarrow A}\cup{\mbb Q}_{\rm E\rightarrow B}$, we define
\alg{
{\rm Qc}({\ca G}):=\log{{\rm dim}Q},
}
where $Q$ is the quantum message transmitted in ${\ca G}$.
The following lemma states that the increment of nMF under quantum communication from Eve to the others is bounded from above by the quantum communication cost. A proof will be given in \rApp{increment}.

\blmm{increment}
For any $\rho\in{\mfk S}_{\rm all}$ and ${\ca G}\in{\mbb Q}_{\rm E}$, it holds that
\alg{
M_F({\ca G}(\rho))-M_F(\rho)\leq{\rm Qc}({\ca G}).
}
\elmm

As discussed in \rSec{NFO}, we have not succeeded in proving that $M_F$ is monotonically nonincreasing under $\Omega^*$. The following lemma states that the $\Omega^*$-monotonicity of $M_F$ is equivalent to the property of ``conditional linearity''.
\blmm{monotnMFC}
The following conditions are equivalent:
\bitem
\setlength{\leftskip}{3mm}
\item[(C1)] $M_F$ is monotonically nonincreasing under $\Omega^*$.
\item[(C2)] $M_F$ is monotonically nonincreasing under $\mbb{C}_{{\rm E}\rightarrow{\rm B}}$.
\item[(C3)] $M_F$ is invariant under $\mbb{C}_{{\rm E}\rightarrow{\rm B}}$.
\item[(C4)]  $M_F$ saturates Inequality \req{convMF}.
\item[(C5)] For any state in the form of \req{decCQst}, it holds that 
 \alg{
\Lambda(\rho)=\sum_mr_m\Lambda(\rho_m).
 \laeq{convLam}
 }
\entem
\elmm

\section{Non-Markovianity Generation}\lsec{disdil}

In this section, 
we  introduce a task in which a non-Markov state is generated from scratch by operations in $\Omega$ and quantum communication from Eve to Alice and Bob, which does not belong to $\Omega$. 
We refer to this task as {\it non-Markovianity generation}. 
We consider an asymptotic limit of infinitely many copies and vanishingly small error. 
The ``non-Markovianity cost'' of a tripartite quantum state is defined as the minimum cost of quantum communication per copy required for non-Markovianity generation.
We prove that the non-Markovianity cost is equal to the regularized non-Markovianity of formation (nMF), by which we provide an operational meaning to nMF.

\subsection{Definitions and Results}
\lsec{dfngeneration}

We first define convertibility of states under operations in $\Omega$. 
Similarly to \rDfn{exactconvertibility}, we take the degree of freedom of reversible operations by Eve into account.
A  rigorous definition is as follows:
\bdfn{econvertibility}
A state $\rho_1$ is {\it $\epsilon$-convertible to $\rho_2$ under $\Omega$} if there exist operations $\ca{F}\in{\Omega}$ and ${\ca V}\in\mbb{R}_{\rm E}$ such that
\alg{
\frac{1}{2}
\left\|\ca{F}(\rho_1^{ABE})-{\ca V}(\rho_2^{ABE})\right\|_1\leq\epsilon,
}
where $\norm{\cdot}_1$ is the trace norm defined by $\norm{A}_1={\rm Tr}|A|$ for an operator $A$.
\edfn

Let $\Omega_Q$ be the set of operations that can be represented as compositions of operations in $\Omega$ and ${\bb Q}_{\rm E}$. 
We define convertibility of states under $\Omega_Q $ analogously to one under $\Omega$, by taking the quantum communication cost from Eve and the others into account. For an operation $\ca{G} \in\Omega_Q $, we quantify the total ``downward'' quantum communication cost by
\alg{
{\rm Qc}(\ca{G}):=
\inf_{\{\ca{F}_k\}_k,\{{\ca G}_k\}_k}
\sum_{k=1}^K{\rm Qc}({\ca G}_k).
}
Here, the infimum is taken over all operations $\{\ca{F}_k\}_k$ and $\{{\ca G}_k\}_k$ such that $\ca{F}_k\in\Omega$, ${\ca G}_k\in{\mbb Q}_{\rm E}$ and
\alg{
\ca{G} ={\ca G}_K\circ\ca{F}_K\circ{\ca G}_{K-1}\circ\ca{F}_{K-1}\circ\cdots\circ{\ca G}_1\circ\ca{F}_1.
\laeq{fsharp}
}
The $\epsilon$-convertibility of states under $\Omega_Q $ is defined as follows:
\bdfn{econvertibility}
A state $\rho_1$ is {\it $\epsilon$-convertible to $\rho_2$ under $\Omega_Q $ with the quantum communication cost $N$} if there exist operations $\ca{G} \in\Omega_Q $ and ${\ca V}\in{\mbb R}_{\rm E}$ such that
\alg{
\frac{1}{2}
\left\|\ca{G} (\rho_1^{ABE})-{\ca V}(\rho_2^{ABE})\right\|_1\leq\epsilon
}
and
\alg{
{\rm Qc}(\ca{G} )\leq N.
}
\edfn

A rigorous definition of the non-Markovianity cost is as follows:
\bdfn{nMcost}
A rate $R$ is {\it achievable in non-Markovianity generation of a state $\rho\in{\mfk S}_{\rm all}$} if, for any $\epsilon>0$ and sufficiently large $n$, the dummy state $\sigma_0:=\proj{0}^A\otimes\proj{0}^B\otimes\proj{0}^E$ is $\epsilon$-convertible to $(\rho^{ABE})^{\otimes n}$ under $\Omega_Q $ with the quantum communication cost $nR$. The {\it non-Markovianity cost} of a state $\rho$, which we denote by $M_C(\rho)$, is defined as the infimum of rate $R$ that is achievable in non-Markovianity generation of $\rho$.
\edfn

The following theorem states that the non-Markovianity cost of a state is equal to the regularized nMF. Proofs are provided in the following subsections.

\bthm{nMcEP}
For any state $\rho$ on $ABE$, it holds that
\alg{
M_C(\rho)= M_F^\infty(\rho).\laeq{wasureta}
}
\ethm

\subsection{Proof of the Converse Part}
\lsec{prfconv}

The converse part of \rThm{nMcEP} is formulated as 
\alg{
M_C(\rho)\geq M_F^\infty(\rho), \laeq{oboeteru3}
}
and is proved as follows. Suppose that a rate $R$ is achievable in non-Markovianity generation of $\rho$. By definition, for any $\epsilon\in(0,1]$ and sufficiently large $n$, there exist operations $\ca{G} \in\Omega_Q$ and ${\ca V}\in\mbb{R}_{\rm E}$ such that 
\alg{
\frac{1}{2}
\left\|\ca{G} (\sigma_0)-\ca{V}((\rho^{ABE})^{\otimes n})\right\|_1\leq\epsilon\laeq{shikaku}
}
and
\alg{
{\rm Qc}(\ca{G})\leq nR.
}
Due to Properties (P8) and (P11) in \rLmm{propIf}, we have
\alg{
M_F(\rho^{\otimes n})=M_F(\ca{V}(\rho^{\otimes n})).
}
and
\alg{
\!\!
M_F(\ca{V}(\rho^{\otimes n}))\leq M_F(\ca{G} (\sigma_0))+4n\sqrt{\epsilon}\log{d_Ad_B}+h'(\epsilon),
\!
}
respectively,
where $h'$ is a nonnegative function that satisfies $\lim_{x\downarrow0}h'(x)=0$. 
In addition, \rLmm{increment} yields
\alg{
M_F(\ca{G} (\sigma_0))\leq {\rm Qc}(\ca{G}).
}
Combining these all together, we arrive at
\alg{
\frac{1}{n}M_F(\rho^{\otimes n})\leq R+4\sqrt{\epsilon}\log{d_Ad_B}+h'(\epsilon).
}
By taking the limit of $n\rightarrow\infty$ and $\epsilon\rightarrow0$, we have $M_F^\infty(\rho)\leq R$. Since this relation holds for any $R>M_C(\rho)$, we obtain \req{oboeteru3}. \QED

\subsection{Proof of the Direct Part}

We prove the direct part of \rThm{nMcEP} by constructing a non-Markovianity generation protocol.
The protocol consists of two steps.
In the first step, an ensemble of pure states is prepared by the quantum reverse Shannon protocol from \cite{abey09}.
In the second step, unnecessary correlation in the state is destroyed by conditional decoupling \cite{berta2018conditional}.
The downward quantum communication costs in the two steps are equal to the two terms in the expression \req{nMF2} of the nMF, respectively.
The details of the protocol are presented as follows.

\subsubsection{Quantum State Redistribution}

\begin{figure}[t]
\begin{center}
\includegraphics[bb={0 0 872 395}, scale=0.29]{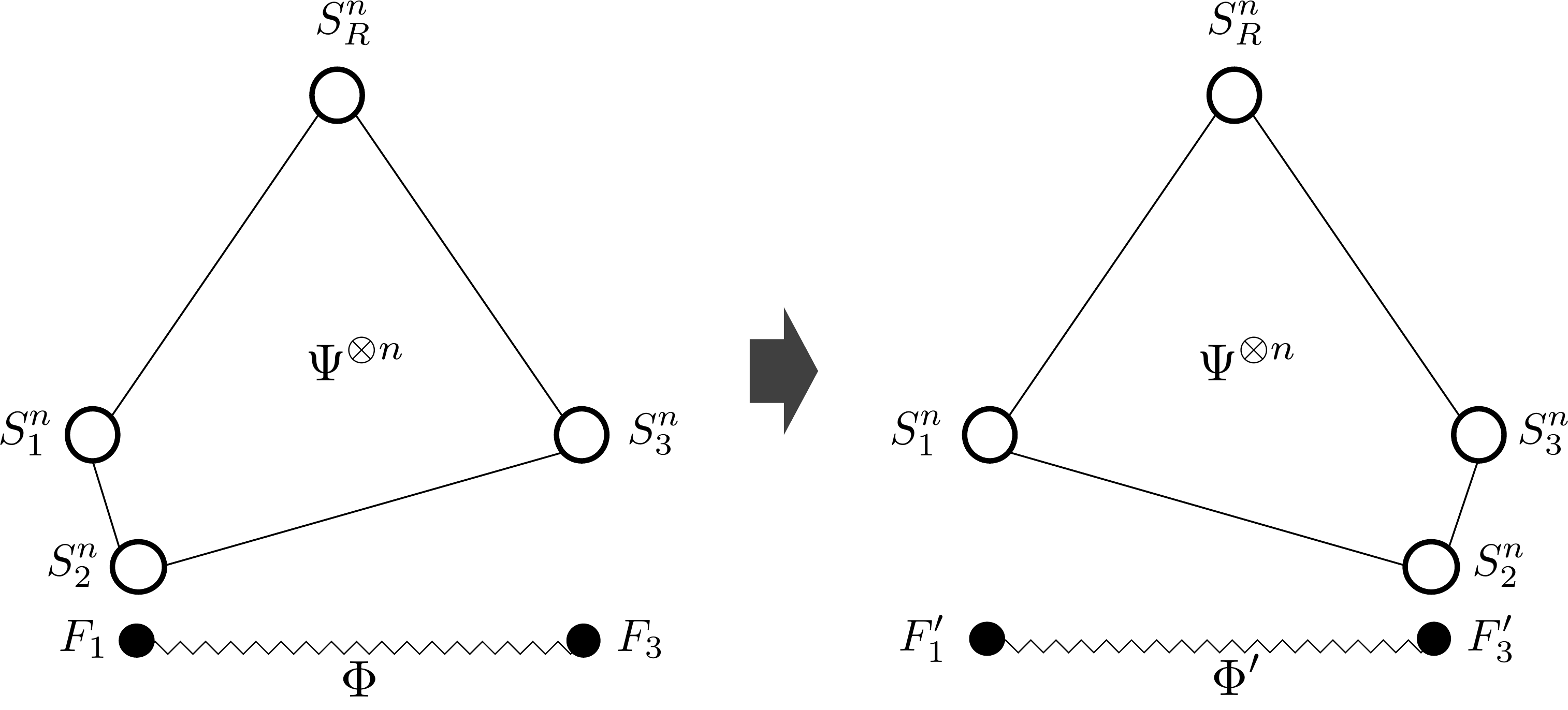}
\end{center}
\caption{The task of quantum state redistribution is depicted. The sender aims at transmitting the $S_2$ part of a pure state $\Psi$ on $S_1S_2S_3S_R$. The systems $S_1$, $S_3$ and $S_R$ are the side information at the sender, the one at the decoder, and the inaccessible reference system, respectively. An entangled state $\Phi$ shared between the sender and the receiver is available as a resource. At the end of the protocol, the sender and the receiver may retrieve some entanglement resource in the form of a state $\Phi'$.}
\label{fig:redisttask01}
\end{figure}

\begin{figure}[h]
\begin{center}
\includegraphics[bb={0 0 327 507}, scale=0.4]{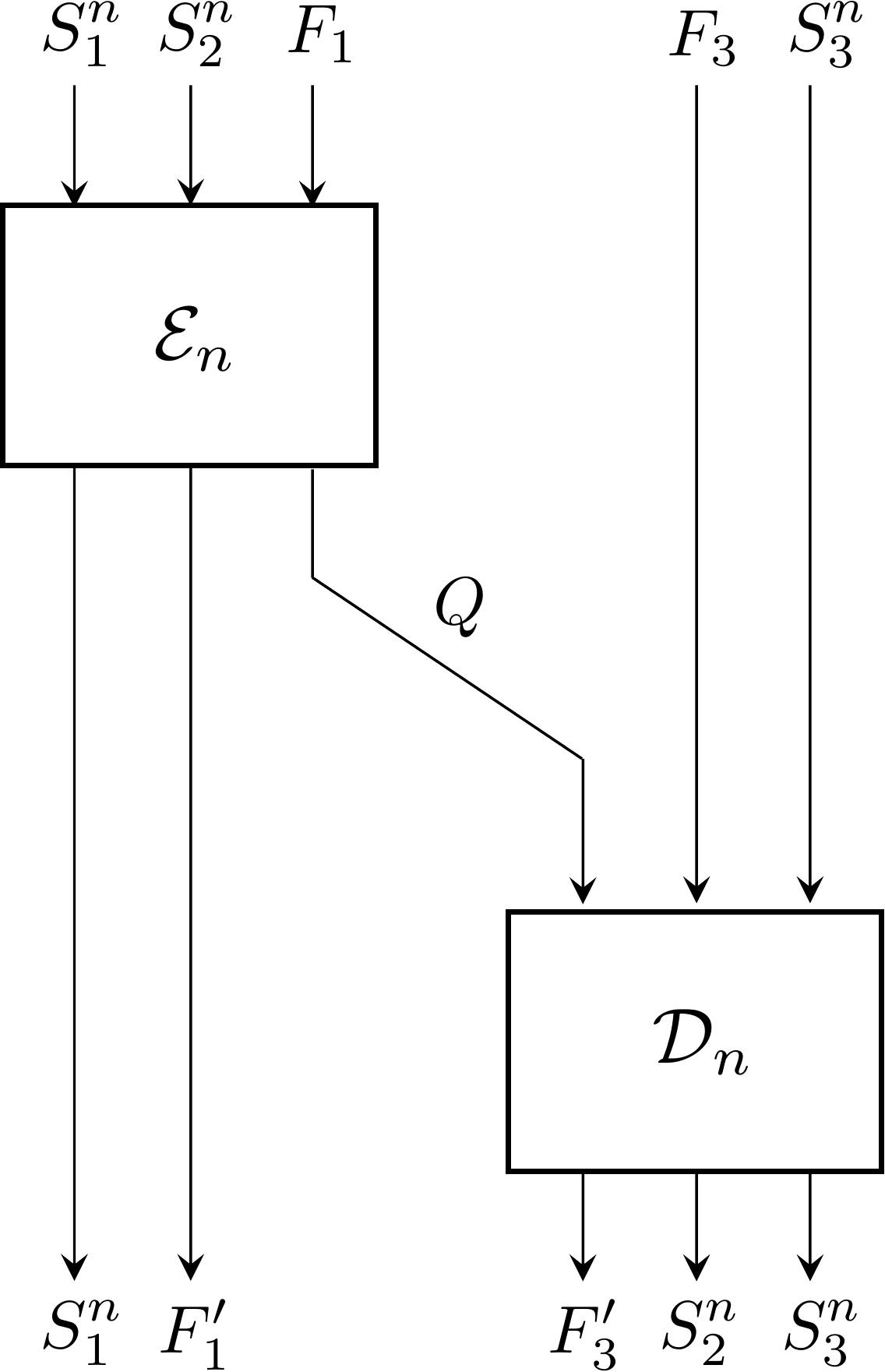}
\end{center}
\caption{A general protocol for quantum state redistribution is depicted. The sender performs an encoding operation $\ca{E}_n$ on the quantum system $S_2$ to be transmitted, the side information $S_1$ and his/her share of the entanglement resource $F_1$. The sender then sends a quantum message $Q$ to the receiver. The receiver subsequently performs a decoding operation $\ca{D}_n$ on the message $Q$, the side information $S_3$ and his/her share of the entanglement resource $F_3$, to obtain the $S_2$ part of the state. }
\label{fig:redisttask02}
\end{figure}

We first review the quantum state redistribution protocol \cite{devetak2008exact,yard2009optimal}, from which the direct part of the quantum reverse Shannon theorem and that of conditional decoupling are obtained by reduction.
Quantum state redistribution is a task in which the $S_2$ part of a four party pure state $\Psi$ on system $S_1S_2S_3S_R$ is transmitted from the sender to the receiver by sending a quantum message $Q$.
Here, $S_R$ is the reference system that is inaccessible to the sender and the receiver, $S_1$ is the side information at the sender and $S_3$ is one at the receiver. 
They may exploit an entangled state $\Phi^{F_1F_3}$ shared in advance as a resource, where $F_1$ and $F_3$ are quantum registers at the sender and the receiver, respectively. 
At the end of the protocol, some part of the entanglement resource may be retrieved in the form of $\Phi'^{F_1'F_3'}$. We are particularly interested in the limit of many copies and vanishingly small error, and in the minimum cost of quantum communication required for the task (see Figure \ref{fig:redisttask01} and \ref{fig:redisttask02}).

For the block length $n$,
the encoding operation and the decoding operation are represented by quantum operations $\ca{E}_n:S_1^nS_2^nF_1\rightarrow S_1^nF_1'Q$ and $\ca{D}_n:S_3^nF_3Q\rightarrow S_2^nS_3^nF_3'$, respectively. The quantum communication cost is quantified by $\log{\dim Q}$. It was proved in \cite{devetak2008exact,yard2009optimal} that the minimum cost of quantum communication is given by the conditional quantum mutual information of the state $\Psi$ (see also \cite{ye2008quantum}). 

\blmm{stateredist}
(Corollary of Theorem 1 in \cite{yard2009optimal}: see also \cite{devetak2008exact})
For any $R>\frac{1}{2}I(S_2:S_R|S_3)_\Psi$, $\epsilon>0$ and sufficiently large $n$, there exist an encoding operation $\ca{E}_n$, a decoding operation $\ca{D}_n$ and entangled states $\Phi^{F_1F_3}$ and $\Phi'^{F_1'F_3'}$ such that
\alg{
&
\frac{1}{2}
\left\|
\ca{D}_n\!\circ\!\ca{E}_n((\Psi^{S_1S_2S_3S_R})^{\otm n}\!\otm\!\Phi^{F_1F_3})
\right.
\nn\\
&
\quad\quad\quad\quad\quad\quad\quad\quad
\left.
-
(\Psi^{S_1S_2S_3S_R})^{\otm n}\!\otm\!\Phi'^{F_1'F_3'}
\right\|_1
\leq\epsilon
\laeq{yakishi}
}
and
\alg{
\log{\dim Q}\leq nR.
}
In addition, we may assume that the encoding operation $\ca{E}_n$ is a linear isometry.
\elmm

The direct part of the quantum reverse Shannon theorem in the version of \cite{abey09} is obtained as a corollary of \rLmm{stateredist} by assuming that $S_3$ is a trivial (one-dimensional) system, i.e., that the receiver has no side information.
By taking the partial trace over $S_2^nS_3^nF_3'$ in \req{yakishi}, we also have
\begin{eqnarray}
\frac{1}{2}
\left\|
{\rm Tr}_{Q}\!\circ\!\ca{E}_n((\Psi^{S_1S_2S_R})^{\otm n}\!\otm\!\Phi^{F_1})
-
(\Psi^{S_1S_R})^{\otm n}\!\otm\!\Phi'^{F_1'}
\right\|_1
\nn\\
\leq\epsilon,
\quad
\laeq{yakishi2}
\end{eqnarray}
which is the direct part of conditional decoupling \cite{berta2018conditional}.

\subsubsection{Direct Part of \rThm{nMcEP} for Pure States}
\lsec{DPPS}

Suppose that $\rho$ is a pure state on $ABE$. Due to (P4) in \rLmm{propIf}, nMF is equal to the quantum mutual information. I.e., for $\rho=|\psi\rangle|\!\langle\psi|^{ABE}$, we have
\alg{
M_F^\infty(\psi)=M_F(\psi)=\frac{1}{2}I(A:B)_\psi.
}
We construct a protocol based on the quantum reverse Shannon theorem.
That is, we apply \rLmm{stateredist} under the following correspondence:
\alg{
S_1\rightarrow E, \: S_2\rightarrow A, \: S_R\rightarrow B, \: \ket{\Psi}\rightarrow\ket{\psi}.
}
We assume that $S_3$ is a trivial system. From \rLmm{stateredist}, it follows that for any $\epsilon>0$, $R>\frac{1}{2}I(A:B|E)_\psi=\frac{1}{2}I(A:B)_\psi$ and sufficiently large $n$, there exist an encoding linear isometry $\ca{V}_n:A^nE^nF_E\rightarrow E^nF_E'Q$, a decoding operation $\ca{D}_n:F_AQ\rightarrow A^nF_A'$  and entangled states $\Phi^{F_EF_A}$ and $\Phi'^{F_E'F_A'}$, such that
\alg{
&
\frac{1}{2}\left\|
\ca{D}_n\!\circ\!\ca{V}_n((\psi^{ABE})^{\otm n}\!\otm\!\Phi^{F_EF_A})
-
(\psi^{ABE})^{\otm n}\!\otm\!\Phi'^{F_E'F_A'}
\right\|_1
\nn\\
&\quad\quad\quad\quad\quad\quad\quad\quad\quad\quad\quad\quad\quad\quad\quad\quad
\quad\quad\quad
\leq\epsilon
\laeq{yakishiii}
}
and that
\alg{
\log{\dim Q}\leq nR.
}

Consider the following protocol $\ca{G}_1$ that is an element of $\Omega_Q$:
\benum
\item Alice locally prepares $\Phi^{F_EF_A}$ and sends the $F_E$ part to Eve.
\item Bob locally prepares $(\ket{\psi}^{ABE})^{\otm n}$ and sends the $A^nE^n$ part to Eve.
\item Eve performs a reversible operation ${\ca V}_n$.
\item Eve sends the quantum system $Q$ to Alice.
\item Alice performs an operation  $\ca{D}_n$ and discards $F_A'$.
\ennum
It is straightforward to verify from \req{yakishiii} that the state obtained by this protocol is equal to $(\psi^{ABE})^{\otimes n}\otm\Phi'^{F_E'}$ up to error $\epsilon$.
The quantum communication cost of the above operation is no greater than $nR$. Hence a rate $R$ is achievable if $R>\frac{1}{2}I(A:B)_\psi$, which implies $M_C(\rho)\leq \frac{1}{2}I(A:B)_\psi$.  \QED

\subsubsection{Direct Part of \rThm{nMcEP} for Mixed States}
\lsec{nMCeqnMF}

Let $\{p_k,\rho_k\}$ be an ensemble of quantum states on $ABE$ satisfying $\rho=\sum_kp_k\rho_k$, and let $\ket{\phi_k}$ be a pure state on $AA'BB'EE'$ such that ${\rm Tr}_{A'B'E'}[\proj{\phi_k}]=\rho_k$ for each $k$. In addition, let $K$ be a quantum system with dimension $|\mathbb{K}|$, and define a state
\alg{
\hat{\rho}:=\sum_{k\in\mathbb{K}}p_k{\rm Tr}_{A'B'}[\proj{\phi_k}]\otimes\proj{k}^{K}.\laeq{dfnrhohat}
}
Fix arbitrary $\epsilon,\delta>0$ and choose sufficiently large $n$. We introduce the following notations:
\alg{
&k^n:=(k_1,\cdots, k_n),\quad p_{k^n}:=p_{k_1}\cdots p_{k_n},\nn\\
&
|\phi_{k^n}\rangle:=\ket{\phi_{k_1}}\cdots\ket{\phi_{k_n}}.\laeq{khkz}
}
We denote by ${\ca T}_{n,\delta}$ be the set of sequences $k^n$ that are $\delta$-strongly typical with respect to $\{p_k\}_k$ (see e.g.~\cite{cover05}).
I.e., ${\ca T}_{n,\delta}$ is the set of all sequences $k^n$ such that
\alg{
|p_k-f_k(k^n)|\leq\delta
\nn
}
for all $k\in\mathbb{K}$,
where $f_k(k^n)$ is the empirical distribution defined by
\alg{
f_k(k^n):=\frac{|\{i|1\leq i\leq n,k_i=k\}|}{n}.
}
Due to the property of the typical set, it holds that
\alg{
\sum_{k^n\in{\ca T}_{n,\delta}}p_{k^n}
\geq
1-\epsilon.
\laeq{wasureyou}
}

Consider the following protocol $\ca{G}_2$ that is an element of $\Omega_Q $:
\benum
\item Alice randomly chooses $k^n$ according to a probability distribution $\{p_{k^n}\}_{k^n}$, and broadcasts it to Bob and Eve.
\item 
If $k^n\notin{\ca T}_{n,\delta}$, Alice, Bob and Eve prepare the dummy state $\sigma_0\in{\mfk S}_{\rm Markov}$.
If $k^n\in{\ca T}_{n,\delta}$, they prepare a state $|\phi_{k^n}'\rangle$, which is equal to $|\phi_{k^n}\rangle$ up to error $\epsilon$, by  the protocol $\ca{G}_1$ presented above. 
\item Alice and Bob discard $A'^n$ and $B'^n$, respectively, and erase $k^n$.
\ennum
This protocol generates a state 
\begin{align}
\hat{\rho}_{n,\delta}=\sum_{k^n\in{\ca T}_{n,\delta}}p_{k^n} \phi_{k^n}'^{\bar{A}\bar{B}\bar{E}\bar{E}'}\otimes\proj{k^n}^{\bar{K}}+p_0\sigma_0,
\end{align}
where
\alg{
p_0:=1-\sum_{k^n\in{\ca T}_{n,\delta}}p_{k_n}\leq\epsilon,\quad
\frac{1}{2}
\left\| \phi_{k^n}'^{\bar{A}\bar{B}\bar{E}\bar{E}'}
-\phi_{k^n}^{\bar{A}\bar{B}\bar{E}\bar{E}'}\right\|_1
\leq\epsilon.
\nn
}
Using \req{wasureyou}, we obtain
\alg{
\frac{1}{2}
\left\|\hat{\rho}_{n,\delta}-\hat{\rho}^{\otimes n}\right\|_1
\leq2\epsilon.
\label{eq:error1}
}
The quantum communication cost of this protocol is equal to one in Step 2. Hence, due to the result in Sec.\ref{sec:DPPS}, we may compose $\ca{G}_2$ so that
\alg{
{\rm Qc}(\ca{G}_2)=\frac{1}{2}n\left(\sum_kp_kI(AA':BB')_{\phi_k}+\delta\right).
\label{eq:coste1}
}

The remaining task is to decouple $E'K$ from the other parts. We adopt the conditional decoupling protocol to address this task.
That is, we apply Inequality \req{yakishi2} by the correspondence:
\alg{
S_1\rightarrow E, \: S_2\rightarrow  E'K, \: R\rightarrow AB, \: \Psi\rightarrow\hat{\rho}.
}
It follows that, for $R=\frac{1}{2}(I(AB:E'K|E)_{\hat{\rho}}+\delta)$, there exists a quantum system $Q$ and a linear isometry $\ca{V}_n:E^nE'^nK^nF_E\rightarrow E^nF_E'Q$ such that
\alg{
&
\frac{1}{2}
\left\|
{\rm Tr}_{Q}\!\circ\!\ca{V}_n((\hat{\rho}^{ABEE'K})^{\otm n}\!\otm\!\Phi^{F_E})
-
(\hat{\rho}^{ABE})^{\otm n}\!\otm\!\Phi'^{F_E'}
\right\|_1
\nn\\
&\quad\quad\quad\quad\quad\quad\quad\quad
\quad\quad\quad\quad\quad\quad\quad\quad
\quad\quad\quad
\leq\epsilon
\laeq{yakishi222}
}
and that $\log{\dim Q}\leq nR$.
Consider the following protocol $\ca{G}_3$ that is an element of $\Omega_Q$:
\benum
\item Eve locally prepares $\Phi^{F_E}$, perform $\ca{V}_n$ and sends quantum system $Q$ to Alice.
\item Alice discards $Q$.
\ennum
The quantum communication cost of $\ca{G}_3$ is bounded as
\alg{
{\rm Qc}(\ca{G}_3)\leq
nR=
\frac{1}{2}n(I(AB:E'K|E)_{\hat{\rho}}+\delta).
\label{eq:coste2}
}

The total error of the protocol is calculated from \req{error1} and \req{yakishi222} as
\alg{
&
\frac{1}{2}
\|\ca{G}_3(\hat{\rho}_{n,\delta})-\rho^{\otimes n}\otm\Phi\|_1
\nn\\
&\leq
\frac{1}{2}
\|\ca{G}_3(\hat{\rho}_{n,\delta})-\ca{G}_3(\hat{\rho}^{\otimes n})\|_1+
\frac{1}{2}
\|\ca{G}_3(\hat{\rho}^{\otimes n})-\rho^{\otimes n}\otm\Phi\|_1
\nn\\
&\leq
\frac{1}{2}
\|\hat{\rho}_{n,\delta}-\hat{\rho}^{\otimes n}\|_1+
\frac{1}{2}
\|\ca{G}_3(\hat{\rho}^{\otimes n})-\rho^{\otimes n}\otm\Phi\|_1\nn\\
&\leq 3\epsilon,
} 
where we have used the fact that $\hat{\rho}^{ABE}=\rho^{ABE}$.
From (\ref{eq:coste1}) and (\ref{eq:coste2}), the total cost of quantum communication is given by
\alg{
&\frac{1}{n}\left({\rm Qc}(\ca{G}_2)+{\rm Qc}(\ca{G}_3)\right)\nn\\
&=\frac{1}{2}\left(\sum_kp_kI(AA':BB')_{\phi_k}+I(AB:E'K|E)_{\hat{\rho}}\right)+\delta.\nn
}
By taking the infimum over all ensembles $\{p_{k},\ket{\phi_k}\}$ and all purifications $\ket{\phi_k}$, the R.H.S. of the above equality is equal to $M_F(\rho)$. Since $\epsilon,\delta>0$ can be arbitrarily small, it follows that a rate $R$ is achievable if $R>M_F(\rho)$. Hence we obtain $M_C(\rho)\leq M_F(\rho)$. By applying this result to the state $\rho^{\otimes m}$, and by taking the limit of $m\rightarrow\infty$, we complete the proof of \rThm{nMcEP}.  
 \hfill$\blacksquare$

\section{The c-Squashed Entanglement}
\lsec{csquashed}

In this section, we apply the result on non-Markovianity generation (\rThm{nMcEP} in \rSec{dfngeneration}) to analyze a measure of bipartite entanglement called the  {\it c-squashed entanglement} (c-SE: \cite{tucci2002entanglement,nagel2003another,yang2009squashed}). c-SE is defined for a state $\omega\in\ca{S}(\ca{H}^{AB})$ by
\alg{
&E_{{\rm sq},c}(\omega^{AB})
:=
\frac{1}{2}
\inf_{\{p_k,\sigma_k\}_{k\in\mbb{K}}}\sum_{k\in\mbb{K}}p_kI(A:B)_{\sigma_k},
\laeq{dfnEsqc}
}
where the infimum is taken over all finite sets $\mbb{K}$ and ensembles $\{p_k,\sigma_k\}_{k\in\mbb{K}}$ of states on $AB$ such that 
\alg{
\sum_{k\in\mbb{K}}p_k\sigma_k^{AB}=\omega^{AB}.
\laeq{magic}
}
We have put the factor $1/2$ in \req{dfnEsqc}, in order that $E_{{\rm sq},c}$ is normalized to $\log{d}$ for a maximally entangled state of Schmidt rank $d$.
c-SE is different from entanglement of formation \cite{bennett96}, in that $\sigma_k$ are not necessarily pure states. The regularized c-squashed entanglement is defined by 
\alg{
E_{{\rm sq},c}^\infty(\omega):=\lim_{n\rightarrow\infty}\frac{1}{n}E_{{\rm sq},c}(\omega^{\otm n}).
}

c-SE satisfies monotonicity under LOCC, convexity, subadditivity, asymptotic continuity and normalization for pure states \cite{yang2009squashed}. However, a direct operational meaning of c-SE is yet unknown.
In the following, we prove that c-SE has a clear operational meaning in terms of a task that we call {\it assisted entanglement dilution}.

\subsection{An Alternative Expression}

We prove that c-SE is represented in terms of the non-Markovianity of formation by ``squashing'' it, in the same way as the squashed entanglement \cite{christandl04} is defined from the conditional quantum mutual information.

\blmm{esqceqmsq}
For any bipartite quantum state $\omega^{AB}$, it holds that
\alg{
E_{{\rm sq},c}(\omega^{AB})
=
M_{{\rm sq}}(\omega^{AB})
:=
\inf_\rho M_F(\rho^{ABE}),
\nn
}
where the infimum is taken over all quantum states $\rho^{ABE}$ such that
\alg{
{\rm Tr}_E[\rho^{ABE}]=\omega^{AB}.
\laeq{oshibori}
}
\elmm

\bprf
First, we prove that $E_{{\rm sq},c}(\omega^{AB})\leq M_{{\rm sq}}(\omega^{AB})$.
Consider an arbitrary state $\rho^{ABE}$ satisfying \req{oshibori}, and let $\{p_k,\ket{\phi_k}\}\in\mbb{K}$ be an ensemble of pure states on $AA'BB'EE'$ that satisfies $\rho^{ABE}=\sum_{k}p_k{\rm Tr}_{A'B'E'}[\proj{\phi_k}]$. 
Defining a state $\varrho^{AA'BB'EE'K}=\sum_{k}p_k\proj{\phi_k}\otm\proj{k}^K$, we have
\alg{
\rho'^{ABK}:={\rm Tr}_{A'B'EE'}[\varrho]=\sum_{k\in\mbb{K}}p_k\phi_k^{AB}\otm\proj{k}^K.
} 
It is straightforward to verify that $\rho'$ is an extension of $\omega^{AB}$. i.e., ${\rm Tr}_K[\rho'^{ABK}]=\omega^{AB}$.
Due to the nonnegativity and monotonicity of  the conditional quantum mutual information under partial trace, we have
$
I(AA':BB'|K)_\varrho+I(AB:E'K|E)_\varrho
\geq
I(A:B|K)_\varrho
=
I(A:B|K)_{\rho'}
\geq
2E_{{\rm sq},c}(\omega^{AB})
$.
By taking the infimum over all finite dimensional quantum systems $A'$, $B'$, $E'$, finite sets $\mbb{K}$ and all ensembles $\{p_k,\ket{\phi_k}\}_{k\in\mbb{K}}$, we obtain
$M_{{\rm sq}}(\omega^{AB})\geq E_{{\rm sq},c}(\omega^{AB})$.

Next, we prove that $E_{{\rm sq},c}(\omega^{AB})\geq M_{{\rm sq}}(\omega^{AB})$.
Let $\{p_k,\sigma_k\}$ be an arbitrary ensemble of states on $AB$ that satisfies $\sum_{k\in\mbb{K}}p_k\sigma_k^{AB}=\omega^{AB}$. 
Let $|\tilde{\phi}_k\rangle^{AB\tilde{E}}$ be an arbitrary purification of $\sigma_k^{AB}$ for each $k$, and define 
$
|\phi_k\rangle^{ABE}:=|\tilde{\phi}_k\rangle^{AB\tilde{E}}\ket{k}^{\tilde{K}}
$,
where $\{\ket{k}\}_{k\in\mbb{K}}$ is an orthonormal set and $E\equiv \tilde{E}\tilde{K}$. Assuming that $A'$, $B'$ and $E'$ are trivial (one-dimensional) systems, the state corresponding to \req{dfnvarrhoK} is given by
\alg{
\varrho^{ABEK}
=
\sum_{k}p_k\proj{\phi_k}^{AB\tilde{E}}
\otm\proj{k}^{\tilde{K}}
\otm\proj{k}^K.
}
Noting that
$
I(AB:K|E)_\varrho=I(AB:K|\tilde{E}\tilde{K})_\varrho=0
$,
we obtain
$
\sum_{k}p_kI(A:B)_{\sigma_k}=I(A:B|K)_\varrho
=I(AA':BB'|K)_\varrho+I(AB:E'K|E)_\varrho
\geq
2M_F(\rho)
$.
By taking the infimum over all ensembles $\{p_k,\sigma_k\}$, 
we obtain $E_{{\rm sq},c}(\omega^{AB})\geq M_{{\rm sq}}(\omega^{AB})$ and completes the proof.
\QED
\eprf

\subsection{Assisted Entanglement Dilution}

\begin{figure}[t]
\begin{center}
\includegraphics[bb={0 60 724 298}, scale=0.33]{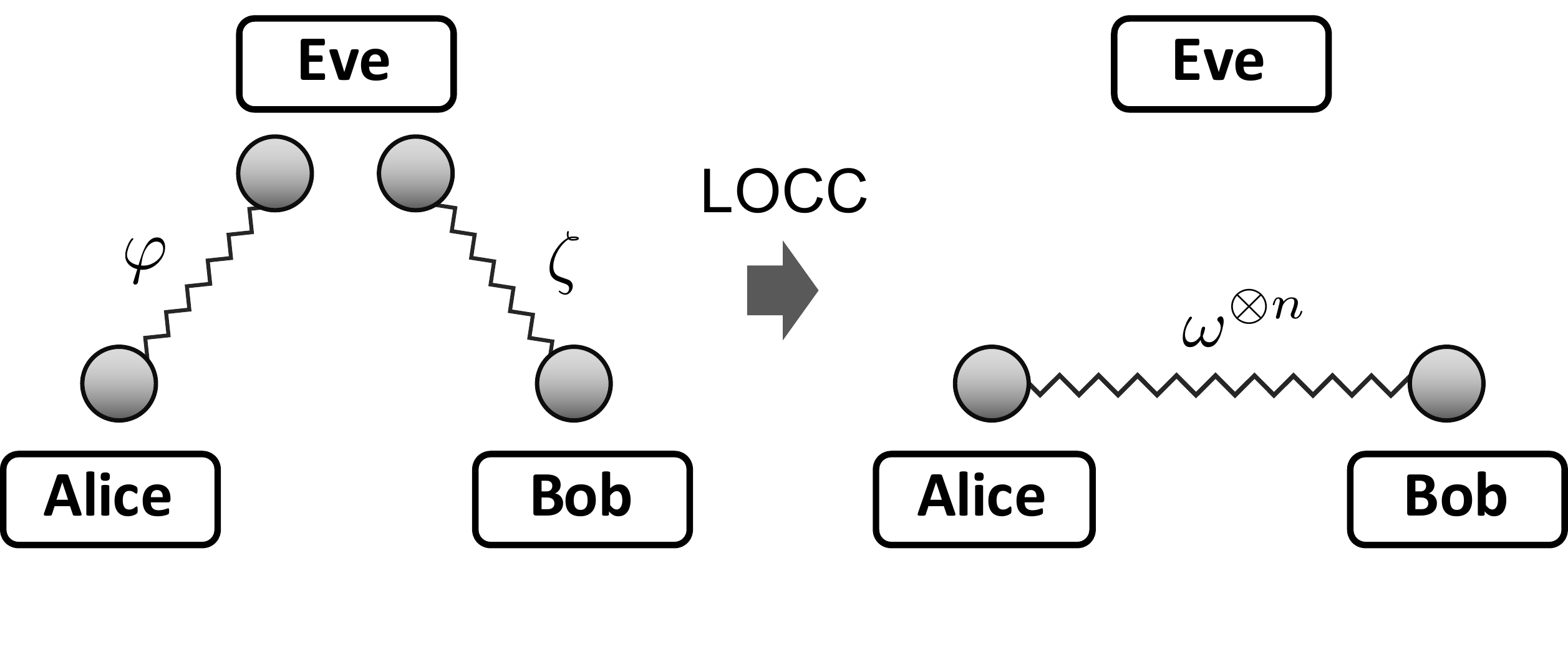}
\end{center}
\caption{Assisted entanglement dilution is depicted. The task is to generate copies of a bipartite quantum state $\omega$ between Alice and Bob by LOCC, with the assistance of Eve who initially shares resource states $\varphi$ and $\zeta$ with Alice and Bob, respectively.}
\label{fig:steering}
\end{figure}

Consider a task in which Alice, Bob and Eve collaborate in order that Alice and Bob share copies of an entangled state $\omega^{AB}$. 
They are allowed to perform arbitrary LOCC protocols. 
In addition, they can exploit bipartite quantum states $\varphi$ and $\zeta$, which are initially shared between Alice and Eve, and between Bob and Eve, respectively. 
We refer to this task as {\it assisted entanglement dilution} (see Figure \ref{fig:steering}).
We focus on a scenario of infinitely many copies and vanishingly small error. 

To achieve this task, it is necessary that Eve performs local measurements and sends the results to the other parties at some point in the protocol.
This is because any dilution protocol that does not involve such an operation induces a state transformation that can be simulated by an LOCC protocol between Alice and Bob.
Note that the reduced state on system $AB$ is initially a product state.
Based on this observation, it would be natural to expect that the degree of entanglement of the target state $\omega^{AB}$ is related to the cost of classical communication from Eve to the other parties, minimized over all dilution protocols for $\omega^{AB}$.
In the following, we prove that this is indeed the case.

Rigorous definitions are as follows.
\begin{dfn}\label{dfn:theprotocol}
Let $\omega^{AB}$ be a bipartite quantum states on system $AB$. Let Alice, Bob and Eve have quantum registers $A_0$, $B_0$ and $\{E_A,E_B\}$, respectively, and let $\ca{T}_n$ be a quantum operation from $A_0\otm B_0\otm E_AE_B$ to ${\bar A}\otm{\bar B}$. $\ca{T}_n$ is called an {\it $(n,\epsilon)$-protocol for generating $\omega^{AB}$}, if $\ca{T}_n$ is an LOCC  and there exist quantum states $\varphi_n^{A_0E_A}$ and $\zeta_n^{B_0E_B}$ such that 
\alg{
\frac{1}{2}
\left\|
\ca{T}_n(\varphi_n^{A_0E_A}\otm\zeta_n^{B_0E_B})
-
(\omega^{AB})^{\otm n}
\right\|_1
\leq
\epsilon.
\laeq{fidelityn}
}
\end{dfn}

Let $\Gamma(\ca{T}_n)$ be the number of times of classical communication from Eve to Alice and Bob in a protocol $\ca{T}_n$.
We denote by $C_{(\gamma)}^{\rm E \rightarrow A}(\ca{T}_n)$ the bit length of  the classical message transmitted in the $\gamma$-th communication from Eve to Alice, and by $C_{(\gamma)}^{\rm E \rightarrow B}(\ca{T}_n)$ that from Eve to Bob, for $\gamma=1,\cdots,\Gamma(\ca{T}_n)$. The ``downward'' classical communication cost of $\ca{T}_n$ is defined as the sum of numbers of classical bits transmitted from Eve to Alice and Bob in $\ca{T}_n$.
That is, we define
\begin{align}
C^\downarrow(\ca{T}_n):=\sum_{\gamma=1}^{\Gamma(\ca{T}_n)}[C_{(\gamma)}^{\rm E \rightarrow A}(\ca{T}_n)+C_{(\gamma)}^{\rm E \rightarrow B}(\ca{T}_n)].\nn
\end{align}
As mentioned above, our interest is on the minimum cost of downward classical communication per copy for accomplishing this task, in an asymptotic limit of $n\rightarrow\infty$ and $\epsilon\rightarrow0$. 

\begin{dfn}\label{dfn:achievablerate}
A rate $R$ is said to be achievable in assisted entanglement dilution of $\omega^{AB}$ if  there exists a sequence of $(n,\epsilon_n)$-protocols $\ca{T}_n$ for generating $\omega^{AB}$ ($n=1,2,\cdots$), with the downward classical communication cost $C^\downarrow(\ca{T}_n)\leq nR$ for each $n$, such that $\lim_{n\rightarrow\infty}\epsilon_n=0$.  
The downward classical communication cost of a state $\omega^{AB}$, which we denote by $C^\downarrow(\omega^{AB})$, is defined as the infimum of rate $R$ that is achievable in assisted entanglement dilution of $\omega^{AB}$.
\end{dfn}

The following theorem states that the downward classical communication cost of a bipartite quantum state is equal to the regularized c-SE. Proofs are provided in the following subsections.

\bthm{moisture}
For any bipartite quantum state $\omega^{AB}$, it holds that
\alg{
C^\downarrow(\omega^{AB})=2E_{{\rm sq},c}^\infty(\omega^{AB}).
}
\ethm

\subsection{Proof of The Converse Part}
\lsec{convAED}

The converse part of \rThm{moisture} is represented as
\alg{
C^\downarrow(\omega^{AB})\geq 2E_{{\rm sq},c}^\infty(\omega^{AB}).
\laeq{convADL}
}
We prove this by using the converse part of non-Markovianity generation (\rThm{nMcEP}).
To this end, we introduce the following lemma, which states that any protocol for assisted entanglement dilution can be converted to a non-Markovianity generation protocol, and that (half of) the downward classical communication cost of the former is equal to the quantum communication cost in the latter. A proof of this lemma will be given in \rApp{prfwawa}.

\blmm{wawawa}
For any $n,l\in\mbb{N}$ and any dilution protocol $\ca{T}_n:A_0\otm B_0\otm E_AE_B\rightarrow A^n\otm B^n$, there exists a quantum system $E_{nl}$ and an operation $\tilde{\ca{T}}_{nl}:A_0^l\otm B_0^l\otm E_A^lE_B^l\rightarrow A^{nl}\otm B^{nl}\otm E_{nl}$, such that $\tilde{\ca{T}}_{nl}\in\Omega_Q$ and
\alg{
&
\ca{T}_n^{\otm l}={\rm Tr}_{E_{nl}}\circ\tilde{\ca{T}}_{nl},
\laeq{hoshishi}\\
&
{\rm Qc}({\tilde{\ca{T}}_{nl}})\leq\left(\frac{l}{2}+1\right)C^\downarrow(\ca{T}_n).
\laeq{hikahika}
}
\elmm

\noindent
{\it Proof of Inequality \req{convADL}:}
By definition, for any $R>C^\downarrow(\omega^{AB})$, $\epsilon>0$, $n,l\in\bb{N}$ and sufficiently large $n$, there exists an $(n,\epsilon/l)$-protocol $\ca{T}_n$ with the downward communication cost $nR$ for assisted entanglement dilution of $\omega^{AB}$. 
That is, for certain states $\varphi_n^{A_0E_A}$ and $\zeta_n^{B_0E_B}$, it holds that
\alg{
\frac{1}{2}
\left\|
\ca{T}_n(\varphi_n^{A_0E_A}\otm\zeta_n^{B_0E_B})
-
(\omega^{AB})^{\otm n}
\right\|_1
\leq
\epsilon/l.
\laeq{masahi}
}
Due to \rLmm{wawawa}, there exists an operation $\tilde{\ca{T}}_{nl}\in\Omega_Q$ from $A_0^l\otm B_0^l\otm E_A^lE_B^l$ to $A^{nl}\otm B^{nl}\otm E_{nl}$ that satisfies Condition \req{hikahika}. 
Consider a state
\alg{
\omega(\tilde{\ca{T}}_{nl})^{A^{nl}B^{nl}E_{nl}}:=
\tilde{\ca{T}}_{nl}((\varphi_n^{A_0E_A})^{\otm l}\otm(\zeta_n^{B_0E_B})^{\otm l}).
\laeq{slasla}
}
From \rLmm{increment} and \req{hikahika}, we have
\alg{
M_F(\omega(\tilde{\ca{T}}_{nl}))
\leq
{\rm Qc}(\tilde{\ca{T}}_{nl})
\leq
\left(\frac{l}{2}+1\right)C^\downarrow(\ca{T}_n).
\laeq{wakama}
}
\rLmm{esqceqmsq} yields
\alg{
M_F(\omega(\tilde{\ca{T}}_{nl}))
&
\geq
M_{{\rm sq}}(\omega(\tilde{\ca{T}}_{nl})^{A^{nl}\!B^{nl}})
\nn\\
&
=
E_{{\rm sq},c}(\omega(\tilde{\ca{T}}_{nl})^{A^{nl}\!B^{nl}}).
\laeq{yukiji}
}
Due to \req{hoshishi} and \req{slasla}, we have
\alg{
\omega(\tilde{\ca{T}}_{nl})^{A^{nl}\!B^{nl}}=
\left(
\ca{T}_n(\varphi^{A_0E_A}\otm\zeta^{B_0E_B})
\right)^{\otm l}.
}
Combining this with \req{masahi}, it follows that
\alg{
\frac{1}{2}
\left\|
\omega(\tilde{\ca{T}}_{nl})^{A^{nl}\!B^{nl}}
-
(\omega^{AB})^{\otm nl}
\right\|_1
\leq
\epsilon.
}
Thus, due to the asymptotic continuity of $E_{{\rm sq},c}$ (see Proposition 5 in \cite{yang2009squashed}), we obtain
\alg{
&
|
E_{{\rm sq},c}(\omega(\tilde{\ca{T}}_{nl})^{A^{nl}\! B^{nl}})
-
E_{{\rm sq},c}(
(\omega^{AB})^{\otm nl}
)
|
\nn\\
&
\quad\quad\quad\quad\quad\quad\quad\quad\quad\quad\quad
\leq
nl\kappa\epsilon
\log{d_Ad_B}
+o(\epsilon).
\laeq{buuta}
}
Here, $\kappa>0$ is a constant and $o(\epsilon)$ is a function satisfying $\lim_{\epsilon\rightarrow0}o(\epsilon)=0$, which are independent of $n$ and the dimensions of the systems.
From \req{wakama}, \req{yukiji} and \req{buuta},
we arrive at
\alg{
&
\frac{1}{nl}
E_{{\rm sq},c}((\omega^{AB})^{\otm nl})
\nn\\
&
\quad\quad
\leq
\left(\frac{1}{2}+\frac{1}{l}\right)
\frac{C^\downarrow(\ca{T}_n)}{n}
+
\kappa\epsilon
\log{d_Ad_B}
+\frac{o(\epsilon)}{nl}.
\nn
}
By taking the limit of $n,l\rightarrow\infty$ and $\epsilon\rightarrow0$, we complete the proof of Ineq.~\req{convADL}.
\QED

\subsection{Proof of The Direct Part}

The direct part of  \rThm{moisture} is represented by
\alg{
C^\downarrow(\omega^{AB})\leq 2E_{{\rm sq},c}^\infty(\omega^{AB}),
\laeq{directAED}
} 
and is proved as follows.
Let $\{p_k,\sigma_k\}$ be an arbitrary ensemble of states on $AB$ that satisfies \req{magic}, and let $|\phi_k\rangle^{ABE}$ be a purification of $\sigma_k^{AB}$ for each $k$. 
We consider a non-Markovianity generation protocol for the state
\alg{
\rho^{ABEK}:=\sum_kp_k\proj{\phi_k}^{ABE}\otm\proj{k}^K.
}
We apply the protocol $\ca{G}_2$ presented in \rSec{nMCeqnMF}, under the assumption that $A'$, $B'$ and $E'$ are trivial (one-dimensional) system. 
For any $\epsilon$ and sufficiently large $n$, the protocol $\ca{G}_2$ generates a state $\hat{\rho}_{n,\delta}^{\bar{A}\bar{B}\bar{E}\bar{K}}$ such that
\alg{
\frac{1}{2}
\left\|
\hat{\rho}_{n,\delta}^{\bar{A}\bar{B}\bar{E}\bar{K}}-(\rho^{ABEK})^{\otm n}
\right\|_1
\leq
2\epsilon.
} 
By tracing out $\bar{E}\bar{K}$, and noting that $\rho^{AB}=\omega^{AB}$, we obtain
\alg{
\frac{1}{2}
\left\|
\hat{\rho}_{n,\delta}^{\bar{A}\bar{B}}-(\omega^{AB})^{\otm n}
\right\|_1
\leq
2\epsilon.
} 
The quantum communication cost of $\ca{G}_2$ is given by \req{coste1}, i.e.,
\alg{
{\rm Qc}(\ca{G}_2)=\frac{1}{2}n\left(\sum_kp_kI(A:B)_{\sigma_k}+\delta\right).
\label{eq:costecosta}
}

We construct an assisted entanglement dilution protocol $\ca{G}_2'$ for $\omega^{AB}$ from $\ca{G}_2$. In $\ca{G}_2'$, instead of communicating quantum messages directly to Alice, Eve performs quantum teleportation by using an entanglement resource shared between Alice and Eve in advance. The downward classical communication cost of this protocol is equal to $n\left(\sum_kp_kI(A:B)_{\sigma_k}+\delta\right)$. By taking the limit of $n\rightarrow\infty$ and $\delta\rightarrow0$, and by taking the infimum over all ensembles $\{p_k,\sigma_k\}$, we obtain 
\alg{
C^\downarrow(\omega^{AB})\leq 2E_{{\rm sq},c}(\omega^{AB}).
}
The same argument also applies to $(\omega^{AB})^{\otm m}$ for any $m\in\mbb{N}$, leading to 
\alg{
C^\downarrow(\omega^{AB})=\frac{1}{m}C^\downarrow((\omega^{AB})^{\otm m})\leq \frac{2}{m}E_{{\rm sq},c}((\omega^{AB})^{\otm m}).
}
By taking the limit of $m\rightarrow\infty$, we arrive at \req{directAED} and completes the proof. 
\QED

\section{Conclusion} \label{sec:discussion}

In this paper, we proposed a framework for analyzing non-Markovianity of tripartite quantum states from the viewpoint of operational resource theory. 
In particular, we introduced a measure of non-Markovianity that we call the non-Markovianity of formation, and provided its operational meaning in terms of a task called non-Markovianity generation. 
Our approach is different from those of \cite{anand2019quantifying,bhattacharya2018convex}, which have investigated non-Markovianity of {\it dynamical processes} in a resource theoretical framework.

It is left open whether the conditional quantum mutual information has a clear operational meaning in the context of the resource theory proposed in this paper.
A future direction is to formulate distillation and dilution of non-Markovianity, and analyze the optimal conversion rate for those tasks. 
It would also be beneficial to address the resource theory of non-Markovianity in a device independent scenario \cite{winczewski2019upper,kaur2020fundamental}.

%
%

\appendices

\section{Properties of The Non-Markovianity of Formation}\lapp{propnMF}

In this Appendix, we prove properties of the non-Markovianity of formation (nMF), presented in \rSec{nMF}. We first prove an alternative expression \req{nMF2} for nMF, and then prove Properties (P1)--(P10) in \rLmm{propIf}, in addition to  \rLmm{increment}. 
A proof of (P11) will be presented in \rApp{asycomnMF}.

\subsection{Proof of Equality \req{nMF2}}
\lapp{prfnMF2}

Define
\alg{
&\hat{M}_F(\rho):=\nn\\
&\frac{1}{2}
\inf_{A',B',E',\mathbb{K}}\inf_{\{p_k,|\phi_k\rangle\}}
\left[I(AA'\!:\!BB'|K)_\varrho+I(AB\!:\!E'K|E)_\varrho\right]\!,\nn
}
where $\varrho$ is the state defined by \req{dfnvarrhoK} and the infimum is taken over all finite dimensional quantum systems $A'$, $B'$, $E'$, finite sets $\mathbb{K}$ and ensembles $\{p_k,|\phi_k\rangle\}_{k\in\mathbb{K}}$ that satisfy \req{decpurho}. 
We prove $M_F(\rho)=\hat{M}_F(\rho)$ by showing that $M_F(\rho)\geq\hat{M}_F(\rho)$ and $M_F(\rho)\leq\hat{M}_F(\rho)$.

To prove the first inequality, let $\{p_k,\rho_k\}_k$ be an arbitrary ensemble of states on $ABE$ such that $\rho=\sum_kp_k\rho_k$. Let $A'$ and $B'$ be finite-dimensional quantum systems, and consider a quantum state  $\tilde{\rho}_k$ on $AA'BB'E$ such that ${\rm Tr}_{A'B'}[\tilde{\rho}_k]=\rho_k$ for each $k$. In addition, let $E'$ be a finite-dimensional quantum system, and let $|\phi_k\rangle^{AA'BB'EE'}$ be a purification of $\tilde{\rho}_k$ for each $k$. 
It is straightforward to verify that the ensemble $\{p_k,|\phi_k\rangle\}_k$ satisfies
\alg{
\rho^{ABE}=\sum_{k}p_k{\rm Tr}_{A'B'E'}[\proj{\phi_k}].
\laeq{deltatoepp}
}
Consider a state $\varrho$ defined by
\alg{
\varrho^{AA'BB'EE'K}=\sum_{k}p_k\proj{\phi_k}^{AA'BB'EE'}\otm\proj{k}^K.
}
We have
\alg{
&\sum_kp_k[S(AA')_{\tilde{\rho}_k}+S(BB')_{\tilde{\rho}_k}]\nn\\
&=\sum_kp_k[S(AA')_{\phi_k}+S(BB')_{\phi_k}]\nn\\
&=S(AA'|K)_\varrho+S(BB'|K)_\varrho\nn\\
&=I(AA':BB'|K)_\varrho+S(AA'BB'|K)_\varrho,\nn
}
in addition to
\alg{
&\sum_kp_kS(A'B')_{\tilde{\rho}_k}
=\sum_kp_kS(A'B')_{\phi_k}
\nn\\
&=\sum_kp_k S(ABEE')_{\phi_k}
\nn\\
&=
\sum_kp_kS(EE')_{\phi_k}+\sum_kp_kS(AB|EE')_{\phi_k}
\nn\\
&=\sum_kp_kS(AA'BB')_{\phi_k}+\sum_kp_kS(AB|EE')_{\phi_k}
\nn\\
&=S(AA'BB'|K)_\varrho+S(AB|EE'K)_\varrho
\laeq{bnio}
}
and
\alg{
S(AB|E)_\varrho-S(AB|EE'K)_\varrho=I(AB:E'K|E)_\varrho.\nn
}
In the second and fourth lines of \req{bnio}, we have used the fact that $|\phi_k\rangle$ is a pure state on $AA'BB'EE'$ for each $k$.
Combining all the above equalities, we obtain
\alg{
&S(AB|E)_\rho+\sum_{k\in\mathbb{K}}p_k\left[S(AA')_{\tilde{\rho}_k}\!+\!S(BB')_{\tilde{\rho}_k}\!-\!S(A'B')_{\tilde{\rho}_k}\right]\nn\\
&=I(AA':BB'|K)_\varrho+I(AB:E'K|E)_\varrho
\geq
2 \hat{M}_F(\rho).\nn
}
By taking the infimum over all finite-dimensional quantum systems $A'$, $B'$, $E'$, all finite sets $\mbb{K}$, all ensembles $\{p_k,\rho_k\}_k$, and over all extensions $\tilde{\rho}_k$ in the L.H.S., we arrive at $M_F(\rho)\geq\hat{M}_F(\rho)$.

The converse inequality $M_F(\rho)\leq\hat{M}_F(\rho)$ is proved along the similar line.
Note that, for any ensemble $\{p_k,|\phi_k\rangle\}_k$ satisfying \req{deltatoepp}, we may define $\tilde{\rho}_k^{AA'BB'E}:={\rm Tr}_{E'}[|\phi_k\rangle\!\langle\phi_k|]$ and $\rho_k^{ABE}:={\rm Tr}_{A'B'}[\tilde{\rho}_k^{AA'BB'E}]$ to obtain $\rho^{ABE}=\sum_kp_k\rho_k^{ABE}$.\\
\QED

\subsection{Proof of (P1): lower bound}

Let $\mathbb{K}$ be a finite set and $\{p_k,\rho_k\}_{k\in\mathbb{K}}$ be an ensemble of states on $ABE$ such that $\rho=\sum_{k\in\mathbb{K}}p_k\rho_k$. Let $A'$, $B'$ and $E'$ be finite dimensional quantum systems.
In addition, let $\tilde{\rho}_k^{AA'BB'E}$ be a state on $AA'BB'E$ satisfying ${\rm Tr}_{A'B'}[\tilde{\rho}_k]=\rho_k^{ABE}$, and let $|\phi_k\rangle$ be a pure state on $AA'BB'EE'$ such that ${\rm Tr}_{A'B'E'}[\proj{\phi_k}]=\rho_k$ for each $k$. 
Let $K$ and $K'$ be quantum systems with dimension $|\mathbb{K}|$, and consider a pure state on $AA'BB'EE'KK'$ defined by
\alg{
|\Phi_\rho\rangle:=\sum_{k\in\mathbb{K}}\sqrt{p_k}|\phi_k\rangle^{AA'BB'EE'}\ket{k}^K\ket{k}^{K'},\nn
}
where $\{\ket{k}\}_k\in\mathbb{K}$ is a set of orthonormal pure states. It is straightforward to verify that the above state is a purification of $\rho^{ABE}$, and that
\alg{
{\rm Tr}_{K'}(\Phi_\rho)=\sum_{k\in\mathbb{K}}p_k\proj{\phi_k}^{AA'BB'EE'}\otimes\proj{k}^K.
}
Consequently, we obtain
\alg{
&\sum_{k\in\mathbb{K}}p_k[S(AA')_{\phi_k}+S(BB')_{\phi_k}]\nn\\
&= S(AA'|K)_{\Phi_\rho}+S(BB'|K)_{\Phi_\rho}
\laeq{CW1}\\
&=S(AA'K)_{\Phi_\rho}+S(BB'K)_{\Phi_\rho}-2S(K)_{\Phi_\rho}
\laeq{CW2}\\
&=S(AA'EE'K')_{\Phi_\rho}+S(BB'EE'K')_{\Phi_\rho}\nn\\
&\quad\quad-2S(AA'BB'EE'K')_{\Phi_\rho}
\laeq{CW3}\\
&=S(AE)_{\Phi_\rho}+S(A'E'K'|AE)_{\Phi_\rho}\nn\\
&\quad\quad+S(BE)_{\Phi_\rho}+S(B'E'K'|BE)_{\Phi_\rho}\nn\\
&\quad\quad-2S(ABE)_{\Phi_\rho}-2S(A'B'E'K'|ABE)_{\Phi_\rho}
\laeq{CW4}\\
&\geq S(AE)_{\Phi_\rho}+S(A'E'K'|ABE)_{\Phi_\rho}\nn\\
&\quad\quad+S(BE)_{\Phi_\rho}+S(B'E'K'|ABE)_{\Phi_\rho}\nn\\
&\quad\quad-2S(ABE)_{\Phi_\rho}-2S(A'B'E'K'|ABE)_{\Phi_\rho}
\laeq{CW5}\\
&= S(AE)_{\Phi_\rho}+S(A'|ABEE'K')_{\Phi_\rho}\nn\\
&\quad\quad+S(BE)_{\Phi_\rho}+S(B'|ABEE'K')_{\Phi_\rho}\nn\\
&\quad\quad-2S(ABE)_{\Phi_\rho}-2S(A'B'|ABEE'K')_{\Phi_\rho}
\laeq{CW6}\\
&=I(A:B|E)_{\Phi_\rho}+S(E)_{\Phi_\rho}+I(A':B'|ABEE'K')_{\Phi_\rho}\nn\\
&\quad\quad-S(ABE)_{\Phi_\rho}-S(A'B'|ABEE'K')_{\Phi_\rho}
\laeq{CW7}\\
&= I(A:B|E)_{\Phi_\rho}-S(AB|E)_{\Phi_\rho}\nn\\
&\quad\quad+I(A':B'|ABEE'K')_{\Phi_\rho}+S(A'B'|K)_{\Phi_\rho}
\laeq{CW8}\\
&\geq I(A:B|E)_{\Phi_\rho}-S(AB|E)_{\Phi_\rho}+S(A'B'|K)_{\Phi_\rho}
\laeq{CW9}\\
&=I(A:B|E)_{\rho}-S(AB|E)_{\rho}+\sum_kp_kS(A'B')_{\phi_k}.
\laeq{CW10}
}
Here, 
\req{CW1} follows from the property of the conditional entropy for classical-quantum states; 
\req{CW2} from the chain rule of the von Neumann entropy; 
\req{CW3} due to the fact that $\Phi_\rho$ is a pure state on $AA'BB'EE'K$; 
\req{CW4} again from the chain rule;
\req{CW5} from the monotonicity of the conditional entropy under discarding of part of the conditioning system;
\req{CW6} from the chain rule; 
\req{CW7} by the definition of the conditional quantum mutual information (CQMI); 
\req{CW8} from the chain rule of the von Neumann entropy and the fact that $\Phi_\rho$ is a pure state on $AA'BB'EE'KK'$; 
\req{CW9} from the nonnegativity of CQMI;
 and \req{CW10} follows from the fact that ${\rm Tr}_{A'B'E'KK'}[\proj{\Phi_\rho}]=\rho$ and the property of the conditional entropy for classical-quantum states.
Noting that $\phi_k^{AA'BB'E}=\tilde{\rho}_k^{AA'BB'E}$,
it follows that
\alg{
&S(AB|E)_\rho+\sum_{k\in\mathbb{K}}p_k[S(AA')_{\tilde{\rho}_k}+S(BB')_{\tilde{\rho}_k}-S(A'B')_{\tilde{\rho}_k}]\nn\\
&\quad\geq I(A:B|E)_{\rho}.
}
By taking the infimum over all $A'$, $B'$, $E'$, $\mathbb{K}$ and $\{p_k,\tilde{\rho}_k^{AA'BB'E}\}_{k\in\mathbb{K}}$ satisfying
\alg{
\sum_{k\in\mathbb{K}}p_k{\rm Tr}_{A'B'}[\tilde{\rho}_k^{AA'BB'E}]=\rho^{ABE},
}
we arrive at $M_F(\rho)\geq \frac{1}{2}I(A:B|E)_{\rho}$. \QED

\subsection{Proof of (P2): upper bound}

Let $A'$ be a trivial (one-dimensional) system, and let $\ket{\phi}$ be an arbitrary pure state on $ABB'E$ such that $\rho^{ABE}={\rm Tr}_{B'}[\proj{\phi}]$. We have
\alg{
&2M_F(A:B|E)_\rho
\nn\\
&\leq
S(AB|E)_\rho+
S(A)_{\phi}+S(BB')_{\phi}-S(B')_{\phi}
\nn\\
&=
S(ABE)_\phi-S(E)_\phi
+
S(A)_{\phi}+S(BB')_{\phi}-S(B')_{\phi}
\nn\\
&=
-S(ABB')_\phi
+
S(A)_{\phi}+S(BB')_{\phi}
\nn\\
&=
I(A:BB')_\phi\leq 2S(A)_{\phi}=2S(A)_{\rho},
\nn
}
where the fourth line follows from the fact that $|\phi\rangle$ is a pure state on $ABB'E$.
Similarly, we also have $M_F(A:B|E)_\rho\leq S(B)_{\rho}$.
\QED

\subsection{Proof of (P3): faithfulness}

The ``only if'' part simply follows from (P1). To prove the ``if'' part, recall that any quantum Markov chain is decomposed in the form of \req{decomposability}. Relabelling $j$ by $k$, define
\alg{
\xi_k:=\Gamma^\dagger(\proj{k}^{E_0}\otm\varsigma_k^{AE_L}\otimes\tau_k^{BE_R})\Gamma.\laeq{mathu}
}
Let $|\psi_{\varsigma_k}\rangle^{AA'E_L}$ and $|\psi_{\tau_k}\rangle^{BB'E_R}$ be purifications of $\varsigma_k^{AE_L}$ and $\tau_k^{AE_R}$, respectively, for each $k$. Define pure states $|\phi_k\rangle$ on $AA'BB'E$ by
\alg{
|\phi_k\rangle:=\Gamma^\dagger\ket{k}^{E_0}|\psi_{\varsigma_k}\rangle^{AA'E_L}|\psi_{\tau_k}\rangle^{BB'E_R}.
}
Consider the state $\varrho$ defined by \req{dfnvarrhoK}, in which we assume that $E'$ is a one-dimensional system. 
It follows that
\alg{
&
\sum_kp_k{\rm Tr}_{A'B'}[\proj{\phi_k}]=\xi,
\nn\\
&
{\rm Tr}_{E}[\proj{\phi_k}]=\psi_{\varsigma_k}^{AA'}\otm\psi_{\tau_k}^{BB'}
\laeq{kashinoki}
}
and that
\alg{
\Gamma\varrho\Gamma^\dagger=\sum_kp_k\proj{\psi_{\varsigma_k}}^{AA'E_L}\otm\proj{\psi_{\tau_k}}^{BB'E_R}\otm\proj{k}^{E_0}\otm\proj{k}^K.
\laeq{kashinokii}
}
From \req{kashinoki},
we have
\alg{
I(AA':BB'|K)_\varrho=\sum_kp_kI(AA':BB')_{\phi_k}=0.
}
In addition, due to \req{kashinokii} and the invariance of CQMI under local isometry, we have
\alg{
I(AB:K|E)_\varrho=I(AB:K|E_0E_LE_R)_{\Gamma(\varrho)}
=0.
}
Combining these equalities with \req{nMF2}, we complete the proof. \\\QED

\subsection{Proof of (P4): pure states}

Let $\ket{\psi}$ be a pure state on $ABE$, and consider an ensemble $\{p_k,\rho_k\}_k$ of states on $ABE$ such that $\proj{\psi}=\sum_kp_k\rho_k$. It is straightforward to verify that $\rho_k=\proj{\psi}$ for all $k$ such that $p_k>0$. 
For each $k$, let $\tilde{\rho}_k$ be a state on $AA'BB'E$ satisfying ${\rm Tr}_{A'B'}[\tilde{\rho}_k]=\rho_k=\proj{\psi}$. It follows that there exists a state $\varsigma_k$ on $A'B'$ such that $\tilde{\rho}_k=|\psi\rangle\!\langle\psi|\otm\varsigma_k$. Hence we have
\alg{
&S(AA')_{\tilde{\rho}_k}+S(BB')_{\tilde{\rho}_k}-S(A'B')_{\tilde{\rho}_k}\nn\\
&=S(A)_{\psi}+S(B)_{\psi}+S(A')_{\varsigma_k}+S(B')_{\varsigma_k}-S(A'B')_{\varsigma_k}\nn\\
&\geq S(A)_{\psi}+S(B)_{\psi},
} 
with equality if and only if $\varsigma_k$ is a product state between $A'$ and $B'$. This implies that  $ \lambda (\rho_k)=S(A)_{\psi}+S(B)_{\psi}$ for all $k$ such that $p_k>0$. Noting that $S(AB|E)_\psi=S(ABE)_\psi-S(E)_\psi=S(AB)_\psi$, we obtain $M_F(\psi)=\frac{1}{2}I(A:B)_\psi$.
\QED

\subsection{Proof of (P5): subadditivity}
\lapp{prfsubadd}

Consider quantum states $\rho$ on $A_1B_1E_1$ and $\sigma$ on $A_2B_2E_2$, and fix ensembles $\{p_k,\rho_k\}_k$ and $\{q_l,\sigma_l\}_l$ such that $\rho=\sum_kp_k\rho_k$ and $\sigma=\sum_lq_l\sigma_l$. 
For each $k$ and $l$, let $\tilde{\rho}_k$ and $\tilde{\sigma}_l$ be states on $A_1A_1'B_1B_1'E_1$ and $A_2A_2'B_2B_2'E_2$, respectively, such that ${\rm Tr}_{A_1'B_1'}[\tilde{\rho}_k]=\rho_k$ and ${\rm Tr}_{A_2'B_2'}[\tilde{\sigma}_l]=\sigma_l$.
We denote $A_1A_2$ by $A$, $A_1'A_2'$ by $A'$, $B_1B_2$ by $B$ and etc.
Noting that $\tilde{\rho}_k\otm\tilde{\sigma}_l$ is an extension of $\rho_k\otm\sigma_l$, we have
\alg{
&S(A_1A_1')_{\tilde{\rho}_k}+S(B_1B_1')_{\tilde{\rho}_k}-S(A_1'B_1')_{\tilde{\rho}_k}\nn\\
&+S(A_2A_2')_{\tilde{\sigma}_l}+S(B_2B_2')_{\tilde{\sigma}_l}-S(A_2'B_2')_{\tilde{\sigma}_l}\nn\\
&=S(AA')_{\tilde{\rho}_k\otm\tilde{\sigma}_l}+S(BB')_{\tilde{\rho}_k\otm\tilde{\sigma}_l}-S(A'B')_{\tilde{\rho}_k\otm\tilde{\sigma}_l}\nn\\
&\geq  \lambda (\rho_k\otm\sigma_l).
}
By taking the infimum over all $\tilde{\rho}_k$, $\tilde{\sigma}_l$, $A_1'B_1'$ and $A_2'B_2'$, we obtain 
\alg{
 \lambda(\rho_k)+ \lambda (\sigma_l)\geq  \lambda (\rho_k\otm\sigma_l).
}
Thus, using the fact that
\alg{
\rho\otimes\sigma=\sum_{k,l}p_kq_l{\rm Tr}_{A'B'}[\tilde{\rho}_k\otm\tilde{\sigma}_l],
}
we arrive at
\alg{
&\Lambda(\rho)+\Lambda(\sigma)\nn\\
&=
\inf_{\mathbb{K}_1}\inf_{\{p_k,\rho_k\}_{k\in\mathbb{K}_1}}\left[\sum_kp_k \lambda (\rho_k)\right]\nn\\
&\quad+\inf_{\mathbb{K}_2}\inf_{\{q_l,\sigma_l\}_{l\in\mathbb{K}_2}}\left[\sum_lq_l \lambda (\sigma_l)\right]
\nn\\
&=
\inf_{\mathbb{K}_1,\mathbb{K}_2}\inf_{\{p_k,\rho_k\}_{k\in\mathbb{K}_1}}\inf_{\{q_l,\sigma_l\}_{l\in\mathbb{K}_2}}
\left[\sum_{k,l}p_kq_l( \lambda (\rho_k)+ \lambda (\sigma_l))\right]
\nn\\
&\geq
\inf_{\mathbb{K}_1,\mathbb{K}_2}\inf_{\{p_k,\rho_k\}_{k\in\mathbb{K}_1}}\inf_{\{q_l,\sigma_l\}_{l\in\mathbb{K}_2}}
\left[\sum_{k,l}p_kq_l \lambda (\rho_k\otm\sigma_l)\right]
\nn\\
&\geq \Lambda(\rho\otm\sigma).
}
Combining this with
\alg{
S(A_1B_1|E_1)_\rho
+S(A_2B_2|E_2)_\sigma
=S(AB|E)_{\rho\otm\sigma},
}
we obtain the desired result.
\QED

\subsection{Proof of (P6): weak chain rule}

Let $\{p_k,|\phi_k\rangle\}_k$ be an ensemble of pure states on $ACA'BB'EE'$ such that $\rho^{ABCE}=\sum_kp_k{\rm Tr}_{A'B'E'}[\proj{\phi_k}]$, and define a state $\varrho$ by
\alg{
&\varrho^{ACA'BB'EE'K}\nn\\
&:=
\sum_{k}p_k\proj{\phi_k}^{ACA'BB'EE'}\otm\proj{k}^K.
}
Due to the chain rule and the monotonicity of CQMI, we have
\alg{
&I(ACA':BB'|K)_{\varrho}+I(ACB:E'K|E)_{\varrho}
\nn\\
&=I(ACA':BB'|K)_{\varrho}+I(C:E'K|E)_{\varrho}
\nn\\
&\quad+I(AB:E'K|EC)_{\varrho}
\nn\\
&\geq
I(AA':BB'|K)_{\varrho}+I(AB:E'K|EC)_{\varrho}
\nn\\
&\geq 2 M_F(A:B|EC)_\rho.
}
By taking the infimum over all $A'$, $B'$, $E'$, $\mathbb{K}$ and $\{p_k,\phi_k\}_{k\in\mathbb{K}}$, we obtain the desired result.
\QED

\subsection{Proof of (P7): conditional convexity} \lapp{DecIfCl}

Let $A'$ and $B'$ be finite-dimensional quantum systems, and let $\mathbb{K}$ be a finite set. For each $m$, let $\{p_{k|m},\tilde{\rho}_{k,m}\}_{k\in\mathbb{K}}$ be an ensemble of states on $AA'BB'E$ such that 
\alg{
\rho_m^{ABE}
=
\sum_{k\in\mathbb{K}}
p_{k|m}
{\rm Tr}_{A'B'}[\tilde{\rho}_{k,m}].
}
Define an ensemble $\{p_{k,m},\tilde{\varsigma}_{k,m}\}_{k\in\mathbb{K},m}$ of states on $AA'BB'EM$ by 
\alg{
\tilde{\varsigma}_{k,m}=\tilde{\rho}_{k,m}^{AA'BB'E}\otm\proj{m}^{M}
\laeq{rhotildek}
}
and
\alg{
p_{k,m}:=p_{k|m}r_m.
\laeq{dfnpkm}
}
It follows that
\alg{
\rho^{ABEM}
=
\sum_{k\in\mathbb{K},m}
p_{k,m}
{\rm Tr}_{A'B'}[\tilde{\varsigma}_{k,m}].
}
Consequently, we obtain
\alg{
&
\sum_{m}r_m\sum_{k\in\mathbb{K}}
p_{k|m} \lambda (\tilde{\rho}_{k,m})
=
\sum_{k\in\mathbb{K},m}
p_{k,m} \lambda (\tilde{\rho}_{k,m})
\nn\\
&=
\sum_{k\in\mathbb{K},m}
p_{k,m} \lambda (\tilde{\varsigma}_{k,m})
\geq
\Lambda(\rho).
\laeq{mouiyada}
}
By taking the infimum over all ensembles $\{p_{k|m},\tilde{\rho}_{k,m}\}_{k\in\mathbb{K}}$ for each $m$, we arrive at
\alg{
\sum_{m}r_m
 \Lambda(\rho_m)
 \geq
  \Lambda(\rho).
}
Combining this with
\alg{
S(AB|EM)_\rho=\sum_mr_mS(AB|E)_{\rho_m},
\laeq{linearCE}
}
we obtain the desired result.
\QED

\subsection{Proof of (P8): invariance under reversible operations}

We first prove that $\Lambda(\rho)$ is monotonically nonincreasing under (possibly irreversible) operations on $E$, that is,
\alg{
\Lambda(\rho)\geq \Lambda(\ca{E}(\rho)),\quad\forall\rho\in\mfk{S}_{\rm all}
,\quad\forall\ca{E}\in\mathbb{L}_E.
\laeq{monotIstae}
}
Let $\{p_k,\rho_k\}_{k\in\mathbb{K}}$ be an ensemble of states on $ABE$ such that $\rho=\sum_kp_k\rho_k$, which yields $\ca{E}(\rho)=\sum_kp_k\ca{E}(\rho_k)$. For each $k$, let $\tilde{\rho}_k^{AA'BB'E}$ be an extension of $\rho_k^{ABE}$.  An extension of $\ca{E}(\rho_k)$ is given by
$
\tilde{\varsigma}_k^{AA'BB'\hat{E}}=\ca{E}(\tilde{\rho}_k^{AA'BB'E})
$.
Noting that $\tilde{\varsigma}_k^{AA'BB'}=\tilde{\rho}_k^{AA'BB'}$,
we have
\alg{
&S(AA')_{\tilde{\rho}_k}+S(BB')_{\tilde{\rho}_k}-S(A'B')_{\tilde{\rho}_k}
\nn\\
&=S(AA')_{\tilde{\varsigma}_k}+S(BB')_{\tilde{\varsigma}_k}-S(A'B')_{\tilde{\varsigma}_k}
\geq  \lambda(\ca{E}(\rho_k)).
}
By taking the infimum over all $A'$, $B'$, $E'$ and $\tilde{\rho}_k$, 
the above inequality yields $ \lambda (\rho_k)\geq  \lambda (\ca{E}(\rho_k))$ . Consequently, we obtain \req{monotIstae} as
\alg{
\Lambda(\rho)
&=\inf_{\mathbb{K}}\inf_{\{p_k,\rho_k\}_{k\in\mathbb{K}}}\left[\sum_{k\in\mathbb{K}}p_k \lambda (\rho_k)\right]
\nn\\
&\geq
\inf_{\mathbb{K}}\inf_{\{p_k,\rho_k\}_{k\in\mathbb{K}}}\left[\sum_{k\in\mathbb{K}}p_k \lambda (\ca{E}(\rho_k))\right]
\nn\\
&\geq
\inf_{\mathbb{K}}\inf_{\{p_k,\rho_k'\}_{k\in\mathbb{K}}}\left[\sum_{k\in\mathbb{K}}p_k \lambda (\rho_k')\right]
=\Lambda(\ca{E}(\rho)),
\nn
}
where the infimum in the last line is taken over all ensembles $\{p_k,\rho_k'\}_{k\in\mathbb{K}}$ such that $\sum_{k\in\mathbb{K}}p_k\rho_k'=\ca{E}(\rho)$.

Let $\ca{V}$ be a reversible operation from $E$ to $\hat{E}$.
It follows from \req{monotIstae} that
\alg{
\Lambda(\rho)\geq \Lambda(\ca{V}(\rho)) 
\geq \Lambda(\ca{V}^*\circ\ca{V}(\rho))= \Lambda(\rho).
} 
Due to the monotonicity of the conditional entropy, we also have
\alg{
&S(AB|E)_\rho\geq S(AB|\hat{E})_{\ca{V}(\rho)} \nn\\
&\geq S(AB|E)_{\ca{V}^*\circ\ca{V}(\rho)}= S(AB|E)_\rho.
} 
Combining these two equalities, we obtain the desired result.
\\\QED

\subsection{Proof of (P9): average monotonicity}

Any measurement $\ca{M}$ on $A$ is represented by
 \alg{
 \ca{M}(\cdot)=\sum_{m}\proj{m}^{A_0}\otm\ca{M}_m(\cdot),
 }
 where $\ca{M}_m$ is a linear CP map from $A$ to $\hat{A}$ such that $\bar{\ca{M}}:=\sum_{m}\ca{M}_m$ is trace-preserving.
 Hence, the state after the measurement is represented by
 \alg{
  \ca{M}(\rho^{ABE})=\sum_{m}\nu_m\proj{m}^{A_0}\otm\rho_m^{\hat{A}BE},
  \laeq{tapioka}
 }
 where we have introduced notations
 \alg{
 \nu_m:={\rm Tr}[\ca{M}_m(\rho)],\quad
 \rho_m^{\hat{A}BE}:=\nu_m^{-1}\ca{M}_m(\rho^{ABE}).
 }
We denote by $V$ a linear isometry from $A$ to $A_0\hat{A}A_e$ such that a Stinespring dilation of $\ca{M}$ is given by $\ca{M}(\cdot)={\rm Tr}_{A_e}[V(\cdot)V^\dagger]$. 

Fix an arbitrary ensemble $\{p_k,|\phi_k\rangle\}_k$ of states on $AA'BB'EE'$ such that $\rho=\sum_kp_k{\rm Tr}_{A'B'E'}[\proj{\phi_k}]$. 
Define 
\alg{
&
p_{m|k}:=\|\bra{m}^{A_0}V\ket{\phi_k}\|^2,
\nn\\
&
\ket{\hat{\phi}_{mk}}:=p_{m|k}^{-1/2}\bra{m}^{A_0}V\ket{\phi_k}.
\nn
}
It is straightforward to verify that
\alg{
\ca{M}(\proj{\phi_k})
=
\sum_mp_{m|k}
\proj{m}^{A_0}
\otm
{\rm Tr}_{A_e}[\proj{\hat{\phi}_{mk}}],
\nn
}
and consequently, that
\alg{
&
\ca{M}(\rho^{ABE})
=
\sum_kp_k{\rm Tr}_{A'B'E'}\circ\ca{M}(\proj{\phi_k})
\nn\\
&
\quad
=
\sum_{k,m}p_kp_{m|k}
\proj{m}^{A_0}
\otm
{\rm Tr}_{A_eA'B'E'}[\proj{\hat{\phi}_{mk}}]
\nn\\
&
\quad
=
\sum_{m}\nu_m
\proj{m}^{A_0}
\otm
\sum_{k}q_{k|m}
{\rm Tr}_{A_eA'B'E'}[\proj{\hat{\phi}_{mk}}]
\nn
}
where we have defined $q_{k|m}:=p_kp_{m|k}/\nu_m$.
Comparing this with \req{tapioka}, we obtain
\alg{
 \rho_m^{\hat{A}BE}
=\sum_{k}q_{k|m}{\rm Tr}_{A_eA'B'E'}[\proj{\hat{\phi}_{mk}}].
\nn
}

Define states $\varrho$, $\hat{\varrho}$ and $\hat{\varrho}_m$ by
\alg{
&
\varrho^{AA'BB'EE'K}
:=
\sum_{k}p_k\proj{\phi_k}^{AA'BB'EE'}
\otm\proj{k}^K,
\nn\\
&
\hat{\varrho}^{A_0\hat{A}A_eA'BB'EE'K}:=V\varrho^{AA'BB'EE'K}V^\dagger
}
and
\alg{
&
\hat{\varrho}_m^{\hat{A}A_eA'BB'EE'K}
\nn\\
&
\quad
:=
\sum_{k}q_{k|m}\proj{\hat{\phi}_{mk}}^{\hat{A}A_eA'BB'EE'}
\otm\proj{k}^K.
}
Denoting by $\ca{D}$ the dephasing operation on $A_0$ with respect to the basis $\{\ket{m}\}$, 
it is straightforward to verify that
\alg{
\ca{D}(\hat{\varrho}^{A_0\hat{A}A_eA'BB'EE'K})
=
\sum_{m}\nu_m
\proj{m}^{A_0}
\otm
\hat{\varrho}_m^{\hat{A}A_eA'BB'EE'K}.
\nn
}
Due to the monotonicity and the chain rule for CQMI, we have
\alg{
&I(AA':BB'|K)_\varrho+I(AB:E'K|E)_\varrho
\nn\\
&=
I(A_0\hat{A}A_eA':BB'|K)_{\hat{\varrho}}
\nn\\
&
\quad\quad
+I(A_0\hat{A}A_eB:E'K|E)_{\hat{\varrho}}\nn\\
&\geq
I(A_0\hat{A}A_eA':BB'|K)_{\ca{D}(\hat{\varrho})}
\nn\\
&
\quad\quad
+I(A_0\hat{A}A_eB:E'K|E)_{\ca{D}(\hat{\varrho})}\nn\\
&=
I(A_0:BB'|K)_{\ca{D}(\hat{\varrho})}
+
I(\hat{A}A_eA':BB'|KA_0)_{\ca{D}(\hat{\varrho})}
\nn\\
&
\quad\quad
+I(A_0:E'K|E)_{\hat{\varrho}}
+I(\hat{A}A_eB:E'K|EA_0)_{\ca{D}(\hat{\varrho})}
\nn\\
&\geq
I(\hat{A}A_eA':BB'|KA_0)_{\ca{D}(\hat{\varrho})}
+I(\hat{A}A_eB:E'K|EA_0)_{\ca{D}(\hat{\varrho})}
\nn\\
&\geq
I(\hat{A}A'':BB'|KA_0)_{\ca{D}(\hat{\varrho})}
\nn\\
&
\quad\quad\quad
+I(\hat{A}B:E'K|EA_0)_{\ca{D}(\hat{\varrho})}
\nn\\
&=
\sum_m\nu_m
I(\hat{A}A'':BB'|K)_{\hat{\varrho}_m}
+
\sum_m\nu_m
I(\hat{A}B:E'K|E)_{\hat{\varrho}_m}
\nn\\
&\geq 
2\sum_m\nu_m M_F(\rho_m),
\nn
}
where we have denoted $A_eA'$ by $A''$ in the ninth line.
By taking the infimum over all $A'$, $B'$, $E'$, $\mathbb{K}$ and $\{p_k,|\phi_k\rangle\}_{k\in\mathbb{K}}$, we obtain the desired result.
\QED

\subsection{Proof of (P10): $\Omega$-monotonicity}
\lapp{monotnMF}

Due to the symmetry of $M_F(A:B|E)$ in $A$ and $B$, 
we only need to prove monotonicity of $M_F$ under ${\bb L}_{\rm A}$, ${\mathbb R}_{\rm E}$, ${\bb Q}_{\rm A\rightarrow E}$ and $\mbb{P}_A$. The first three cases directly follow from (P9), (P8) and (P6), respectively. 
To prove monotonicity under $\mbb{P}_A$, note that the state before and after broadcasting of classical message by Alice is represented by density operators
\alg{
\rho_i=\sum_mr_m\proj{m}^{M_{\!A}}\otimes\rho_m^{ABE}
} 
and
\alg{
\rho_f=\sum_mr_m\proj{m}^{M_{\!A}}\otimes\proj{m}^{M_{\!B}}\otimes\proj{m}^{M_{\!E}}\otimes\rho_m^{ABE},
} 
respectively.
Due to the average monotonicity (P9) and the conditional convexity (P7), it follows that
\alg{
M_F(AM_A:B|E)_{\rho_i}
&
\geq
\sum_mr_m
M_F(A:B|E)_{\rho_m}
\nn\\
&\geq
M_F(AM_A:BM_B|EM_E)_{\rho},
}
which completes the proof.
\QED

\subsection{Proof of \rLmm{increment} (Notations modified $C\rightarrow Q$.)}
\lapp{increment}

Due to the symmetry of $M_F(A:B|E)$ in $A$ and $B$, 
we only need to consider $\mbb{Q}_{\rm E\rightarrow A}$.
Let $Q$ be the quantum system that is transmitted from Eve to Alice.
Consider the state $\rho^{ABQE}$, and
let $\{p_k,\tilde{\rho}_k\}$ be an arbitrary ensemble of states on $AA'BB'EQ$ such that $\rho^{ABQE}=\sum_k p_k{\rm Tr}_{A'B'}[\tilde{\rho}_k]$. We have
\alg{
&2M_F(AQ:B|E)_\rho
\nn\\
&\leq
S(AQB|E)_\rho+\nn\\
&\quad\quad
\sum_kp_k[S(AQA')_{\tilde{\rho}_k}+S(BB')_{\tilde{\rho}_k}-S(A'B')_{\tilde{\rho}_k}]
\nn\\
&=
S(AB|EQ)_\rho+\nn\\
&\quad\quad
\sum_kp_k[S(AA')_{\tilde{\rho}_k}+S(BB')_{\tilde{\rho}_k}-S(A'B')_{\tilde{\rho}_k}]
\nn\\
&\quad\quad+S(Q|E)_\rho+\sum_kp_kS(Q|AA')_{\tilde{\rho}_k}.
\laeq{hukuhuku}
}
Due to the monotonicity and the concavity of the conditional quantum entropy, we have
\alg{
\sum_kp_kS(Q|AA')_{\tilde{\rho}_k}
\leq
\sum_kp_kS(Q|A)_{\tilde{\rho}_k}
\leq
S(Q|A)_{\rho}.
}
Substituting this to \req{hukuhuku}, and noting that $S(Q|E)_\rho+S(Q|A)_{\rho}\leq 2S(Q)\leq 2\log{d_Q}$, we obtain
\alg{
&2M_F(AQ:B|E)_\rho
\nn\\
&\leq
S(AB|EQ)_\rho+\nn\\
&\quad
\sum_kp_k[S(AA')_{\tilde{\rho}_k}+S(BB')_{\tilde{\rho}_k}-S(A'B')_{\tilde{\rho}_k}]
+2\log{d_Q}.
\nn
}
By taking the infimum over all $A'$, $B'$ and $\{p_k,\tilde{\rho}_k\}$, we obtain the desired result.
\QED

\subsection{Proof of \rLmm{monotnMFC}}

(C1)$\Rightarrow$(C2) follows from the definition of $\Omega^*$.
(C2)$\Rightarrow$(C1) follows from the $\Omega$-monotonicity of $M_F$ and symmetry of $M_F(A:B|E)$ in $A$ and $B$.
(C3)$\Rightarrow$(C2) immediately follows by definition.
Equivalence between (C4) and (C5) follows from \req{dfnIf} and \req{linearCE}. 

We compete the proof by showing that (C2)$\Rightarrow$(C4)$\Rightarrow$(C3).
Recall that the states before and after classical communication from Eve to Bob is represented by density operators \req{rhohati2} and \req{rhohatf3}, respectively. 
Suppose that $M_F$ is monotonically nonincreasing under $\mbb{C}_{\rm E\rightarrow B}$, i.e., $M_F(\rho_i) \geq M_F(\rho_f)$. Due to the conditional convexity (P7), we have 
\alg{
&
\sum_mr_mM_F(A:B|E)_{\rho_m}
\geq
M_F(\rho_i)
\geq
M_F(\rho_f).
\nn
}
In addition, due to the average monotonicity (P9) under the measurement on $M_B$ with respect to the basis $\{|m\rangle\}_m$, we have 
\alg{
M_F(\rho_f)
\geq
\sum_mr_mM_F(A:B|E)_{\rho_m}.
}
Combining these two inequalities, we obtain $M_F(\rho_i)=\sum_mr_mM_F(A:B|E)_{\rho_m}$, which implies (C2)$\Rightarrow$(C4).
 Next, suppose that $M_F$ saturates Inequality \req{convMF},
in which case we have
\alg{
M_F(\rho_i)
=\sum_mr_mM_F(A:B|E)_{\rho_m}
=M_F(\rho_f).
}
This implies (C4)$\Rightarrow$(C3) and completes the proof.
\QED

\section{Asymptotic Continuity of The non-Markovianity of Formation}\lapp{asycomnMF}

In this appendix, we prove Property (P11) in \rLmm{propIf}, i.e., asymptotic continuity of the non-Markovianity of formation. The proof proceeds along the same line as the proof of asymptotic continuity of entanglement of formation, in the version of Corollary 4 in \cite{winter2016tight}.

\bthm{asyconIf}
For any states $\rho$ and $\sigma$ on system $ABE$ such that $\frac{1}{2}\|\rho-\sigma\|_1\leq\epsilon\leq1$, it holds that 
\begin{align}
\left|M_F(\rho)-M_F(\sigma)\right|\leq
4\sqrt{\epsilon}\log{d_Ad_B}+3(1+\sqrt{\epsilon})h\left(\frac{\sqrt{\epsilon}}{1+\sqrt{\epsilon}}\right),
\laeq{yosoyoso}
\end{align}
where $h$ is the binary entropy defined by $h(x)=-x\log{x}-(1-x)\log{(1-x)}$.
\ethm

\bprf
Suppose that $\frac{1}{2}\|\rho-\sigma\|_1\leq\epsilon\leq1$. 
The Alicki-Fannes inequality (\!\!\cite{alicki04}: see \cite{winter2016tight} for an improved version) yields
\alg{
|
S(AB|E)_{\sigma}
-
S(AB|E)_{\rho}
|
\leq
2\epsilon\log{d_Ad_B}+(1+\epsilon)h\left(\frac{\epsilon}{1+\epsilon}\right).
\nn
}
As we prove below, it also holds that
\alg{
|\Lambda(\rho)
-
\Lambda(\sigma)|
\leq
2\sqrt{\epsilon}\log{d_Ad_B}+2(1+\sqrt{\epsilon})h\left(\frac{\sqrt{\epsilon}}{1+\sqrt{\epsilon}}\right).
\laeq{rashirashi}
}
Combining these two inequalities, we obtain \req{yosoyoso}.

To prove \req{rashirashi}, we may, without loss of generality, assume that
\alg{
\Lambda(\sigma)\leq \Lambda(\rho).
\laeq{sagashita}
} 
Due to the condition $\frac{1}{2}\|\rho-\sigma\|_1\leq\epsilon$ and Proposition 5 in \cite{winter2016tight}, there exists a purification $|\psi_\sigma\rangle^{ABER}$ of $\sigma$ and a state $\theta_\rho^{ABER}$ such that
\alg{
\frac{1}{2}
\||\psi_\sigma\rangle\!\langle\psi_\sigma|-\theta_\rho\|_1\leq\sqrt{\epsilon}\laeq{sulver}
}
and that
\alg{
&
\theta_\rho^{ABE}=\rho^{ABE},
\laeq{bokutachi}
\\
&
\theta_\rho^R=\psi_\sigma^R.
\laeq{chitteyukukara}
}

Let $A'$ and $B'$ be finite dimensional quantum systems, $\mathbb{K}$ be a finite set and $\{p_k,\tilde{\sigma}_k\}_{k\in\mathbb{K}}$ be an ensemble of states on $AA'BB'E$ such that 
\alg{
\sigma^{ABE}=\sum_{k\in\mbb{K}}p_k{\rm Tr}_{A'B'}[\tilde{\sigma}_k].
}
There exists a quantum operation $\ca{M}:R\rightarrow A'B'K$, which is in the form of
\alg{
\ca{M}(\cdot)=\sum_{k\in\mathbb{K}}M_k(\cdot)M_k^\dagger \otm\proj{k}^K,
\quad
\inpro{k}{k'}=\delta_{k,k'},
\nn
}
 such that
\alg{
\tilde{\sigma}:=\ca{M}(\psi_\sigma)
=\sum_{k\in\mathbb{K}}p_k\tilde{\sigma}_k^{AA'BB'E}\otm\proj{k}^K.
\laeq{paipai}
}
It holds that
\alg{
&
\sum_{k\in\mathbb{K}}p_k[S(AA')_{\tilde{\sigma}_k}+S(BB')_{\tilde{\sigma}_k}-S(A'B')_{\tilde{\sigma}_k}]
\nn\\
&=
S(AA'|K)_{\tilde{\sigma}}+S(BB'|K)_{\tilde{\sigma}}-S(A'B'|K)_{\tilde{\sigma}}
\nn\\
&=
S(A|A'K)_{\tilde{\sigma}}+S(B|B'K)_{\tilde{\sigma}}+I(A':B'|K)_{\tilde{\sigma}}
.
\laeq{kagirias}
}
Applying the same map $\ca{M}$ to the state $\theta_\rho$, we obtain
\alg{
\tilde{\rho}:=\ca{M}(\theta_\rho)
=\sum_{k\in\mathbb{K}}p_k\tilde{\rho}_k^{AA'BB'E}\otm\proj{k}^K,
\laeq{paipaipai}
}
where
\alg{
\tilde{\rho}_k^{AA'BB'E}:=
p_k^{-1}M_k(\theta_\rho)M_k^\dagger
.
}
Note that, due to \req{chitteyukukara}, it holds that
\alg{
{\rm Tr}[\tilde{\rho}_k]=p_k^{-1}{\rm Tr}[M_k(\theta_\rho)M_k^\dagger]=p_k^{-1}{\rm Tr}[M_k(\psi_\sigma)M_k^\dagger]=1.
\nn
}
In addition, from \req{bokutachi}, it follows that
\alg{
\sum_{k\in\mathbb{K}}p_k\tilde{\rho}_k^{ABE}
=
{\rm Tr}_{A'B'K}\circ\ca{M}(\theta_\rho)
=
\theta_\rho^{ABE}
=
\rho^{ABE}.
\nn
}
Thus, similarly to \req{kagirias}, we have
\alg{
\Lambda(\rho)
&
\leq
\sum_{k\in\mathbb{K}}p_k[S(AA')_{\tilde{\rho}_k}+S(BB')_{\tilde{\rho}_k}-S(A'B')_{\tilde{\rho}_k}]
\nn\\
&=
S(A|A'K)_{\tilde{\rho}}+S(B|B'K)_{\tilde{\rho}}+I(A':B'|K)_{\tilde{\rho}}.
\laeq{kagirias2}
}
Since \req{chitteyukukara} implies $\tilde{\sigma}^{A'B'K}=\tilde{\rho}^{A'B'K}$,
we also have
\alg{
I(A':B'|K)_{\tilde{\sigma}}
=
I(A':B'|K)_{\tilde{\rho}}.
\laeq{kehai}
}
Combining \req{kagirias}, \req{kagirias2} and \req{kehai}, we arrive at
\alg{
\Lambda(\rho)
&
\leq
\sum_{k\in\mathbb{K}}p_k[S(AA')_{\tilde{\sigma}_k}+S(BB')_{\tilde{\sigma}_k}-S(A'B')_{\tilde{\sigma}_k}]
\nn\\
&\quad\quad
+S(A|A'K)_{\tilde{\rho}}-S(A|A'K)_{\tilde{\sigma}}
\nn\\
&\quad\quad
+S(B|B'K)_{\tilde{\rho}}-S(B|B'K)_{\tilde{\sigma}}.
\laeq{kehai3}
}

From \req{sulver}, \req{paipai}, \req{paipaipai} and the monotonicity of the trace distance, we have
\alg{
\frac{1}{2}
\left\|\tilde{\rho}-\tilde{\sigma}\right\|_1
\leq
\frac{1}{2}
\left\|\theta_\rho-\proj{\psi_\sigma}\right\|_1
\leq\sqrt{\epsilon}.
\nn
}
Consequently, the Alicki-Fannes inequality yields
\alg{
&
|S(A|A'K)_{\tilde{\rho}}-S(A|A'K)_{\tilde{\sigma}}|
\nn\\
&\quad\quad
\leq
2\sqrt{\epsilon}\log{d_A}+(1+\sqrt{\epsilon})h\left(\frac{\sqrt{\epsilon}}{1+\sqrt{\epsilon}}\right),
\laeq{kanas}
\\
&
|S(B|B'K)_{\tilde{\rho}}-S(B|B'K)_{\tilde{\sigma}}|
\nn\\
&\quad\quad
\leq
2\sqrt{\epsilon}\log{d_B}+(1+\sqrt{\epsilon})h\left(\frac{\sqrt{\epsilon}}{1+\sqrt{\epsilon}}\right).
\!
\laeq{kanas2}
}
Hence, from \req{kehai3}, we arrive at
\alg{
\Lambda(\rho)
&
\leq
\sum_{k\in\mathbb{K}}p_k[S(AA')_{\tilde{\sigma}_k}+S(BB')_{\tilde{\sigma}_k}-S(A'B')_{\tilde{\sigma}_k}]
\nn\\
&\quad\quad\quad
+
2\sqrt{\epsilon}\log{d_Ad_B}+2(1+\sqrt{\epsilon})h\left(\frac{\sqrt{\epsilon}}{1+\sqrt{\epsilon}}\right).
\nn
}
Taking the infimum over all $A'$, $B'$, $\mathbb{K}$ and $\{p_k,\tilde{\sigma}_k\}_{k\in\mathbb{K}}$, the above inequality yields
\alg{
\Lambda(\rho)
\leq
\Lambda(\sigma)
+
2\sqrt{\epsilon}\log{d_Ad_B}+2(1+\sqrt{\epsilon})h\left(\frac{\sqrt{\epsilon}}{1+\sqrt{\epsilon}}\right).
\nn
}
Combining this with \req{sagashita}, we obtain \req{rashirashi}
and complete the proof. \QED
\eprf

\section{Proof of \rLmm{wawawa}}
\lapp{prfwawa}

\begin{figure}[t]
\begin{center}
\includegraphics[bb={0 20 421 235}, scale=0.35]{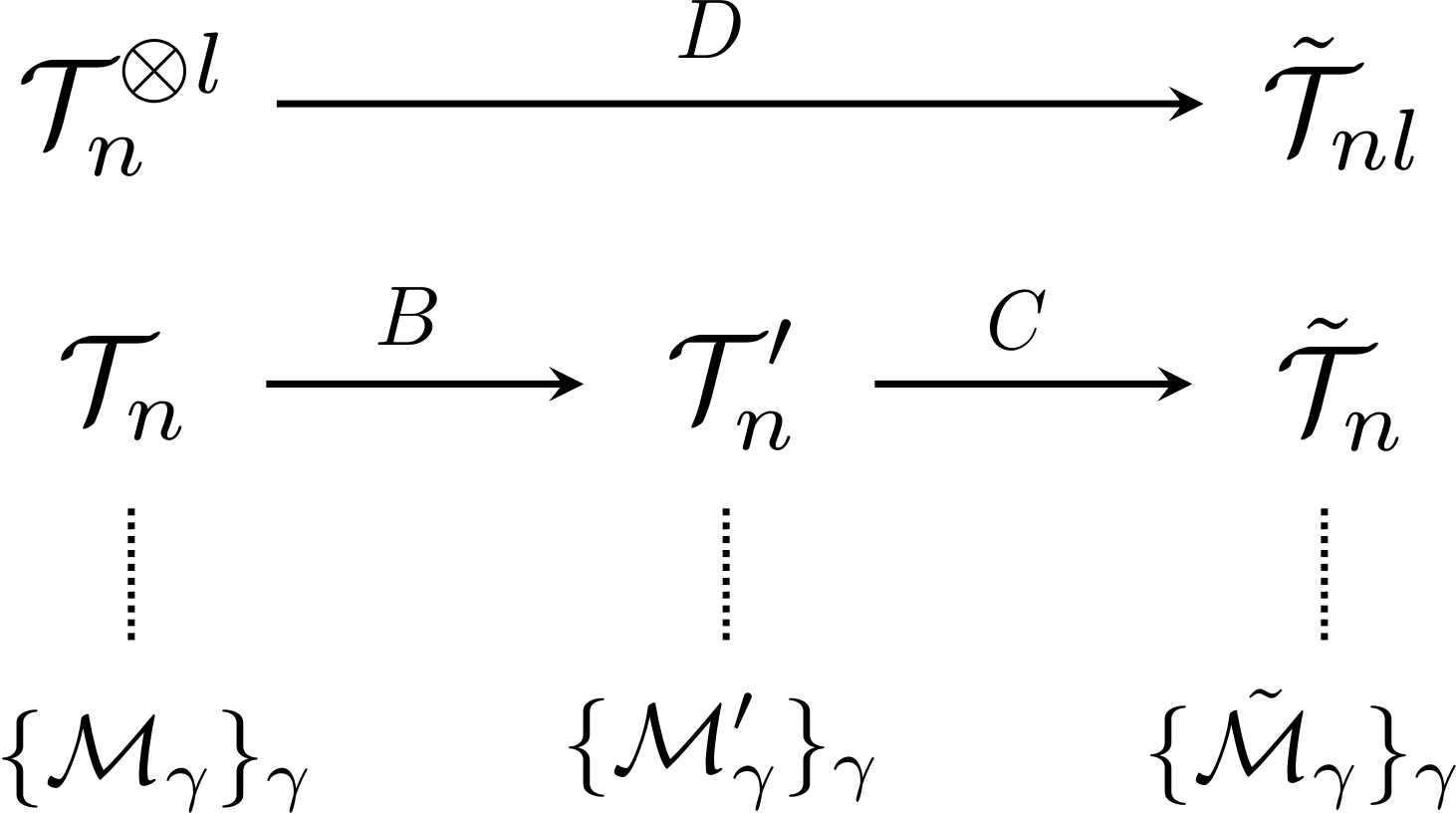}
\end{center}\caption{Construction of a protocol $\tilde{\ca{T}}_{nl}$ from an assisted entanglement dilution protocol $\ca{T}_n^{\otm l}$, satisfying the conditions described in \rLmm{wawawa}, is depicted. The alphabets over the arrows point to subsections in \rApp{prfwawa} in which the conversion of protocols are presented.}
\label{fig:PRTCVS}
\end{figure}

In this Appendix, we prove \rLmm{wawawa} in \rSec{convAED}.
The lemma states that any protocol for assisted entanglement dilution can be converted to a non-Markovianity generation protocol, and that the downward classical communication cost of the former is equal to half of the quantum communication cost in the latter.
To this end, we first introduce a description of an arbitrary dilution protocol. 
In particular, we investigate the description of operations by Eve in detail.
Based on the obtained description, we construct a non-Markovianity generation protocol from an assisted entanglement dilution protocol, such that the communication cost satisfies the above condition (see Figure \ref{fig:PRTCVS}).

\subsection{Description of Dilution Protocols}

Without loss of generality, we may assume that any dilution protocol proceeds as follows.
Here, $\Gamma$ is the number of communication rounds in the protocol.
Symbols $K_\gamma$, $L_\gamma$, $M_\gamma$ and $\hat{M}_\gamma$ are for random variables that represent the classical messages communicated among the parties and the outcomes of the measurement by Eve, respectively.
 We use the same symbols for systems in which those variables are registered.
These variables take values in certain finite sets ${\mfk K}_\gamma$, ${\mfk L}_\gamma$, ${\mfk M}_\gamma$ and $\hat{\mfk M}_\gamma$, respectively. 
\begin{enumerate}
\item Alice, Bob and Eve recursively apply the following operation from $\gamma=1$ to $\gamma=\Gamma$: 
\begin{enumerate}
\item Alice performs a measurement on her system and obtains an outcome.
\item Alice broadcasts a classical message $K_\gamma$ to Bob and Eve.
\item Bob performs a measurement on his system and obtains an outcome.
\item Bob broadcasts a classical message $L_\gamma$ to Alice and Eve.
\item Eve performs a measurement ${\mathcal M}_\gamma$ on her system and obtains an outcome $\hat{M}_\gamma$.
\item Eve broadcasts a classical message $M_\gamma$ to Alice and Bob.
\end{enumerate}
\item Alice and Bob perform local operations on their systems.
\item Eve discards all of her systems.
\end{enumerate}
Denoting the cardinality of ${\mfk M}_\gamma$ by $\mu_\gamma$,
the total number of classical bits, broadcasted by Eve from Alice to Bob during the protocol, is given by
\begin{align}
C^\downarrow(\ca{T}_n):=\sum_{\gamma=1}^\Gamma\log{\mu_\gamma}.\nn
\end{align}
It should be noted that all operations in Step (a)-(d) above belong to $\Omega$.

\begin{figure}[t]
\begin{center}
\includegraphics[bb={0 20 462 466}, scale=0.35]{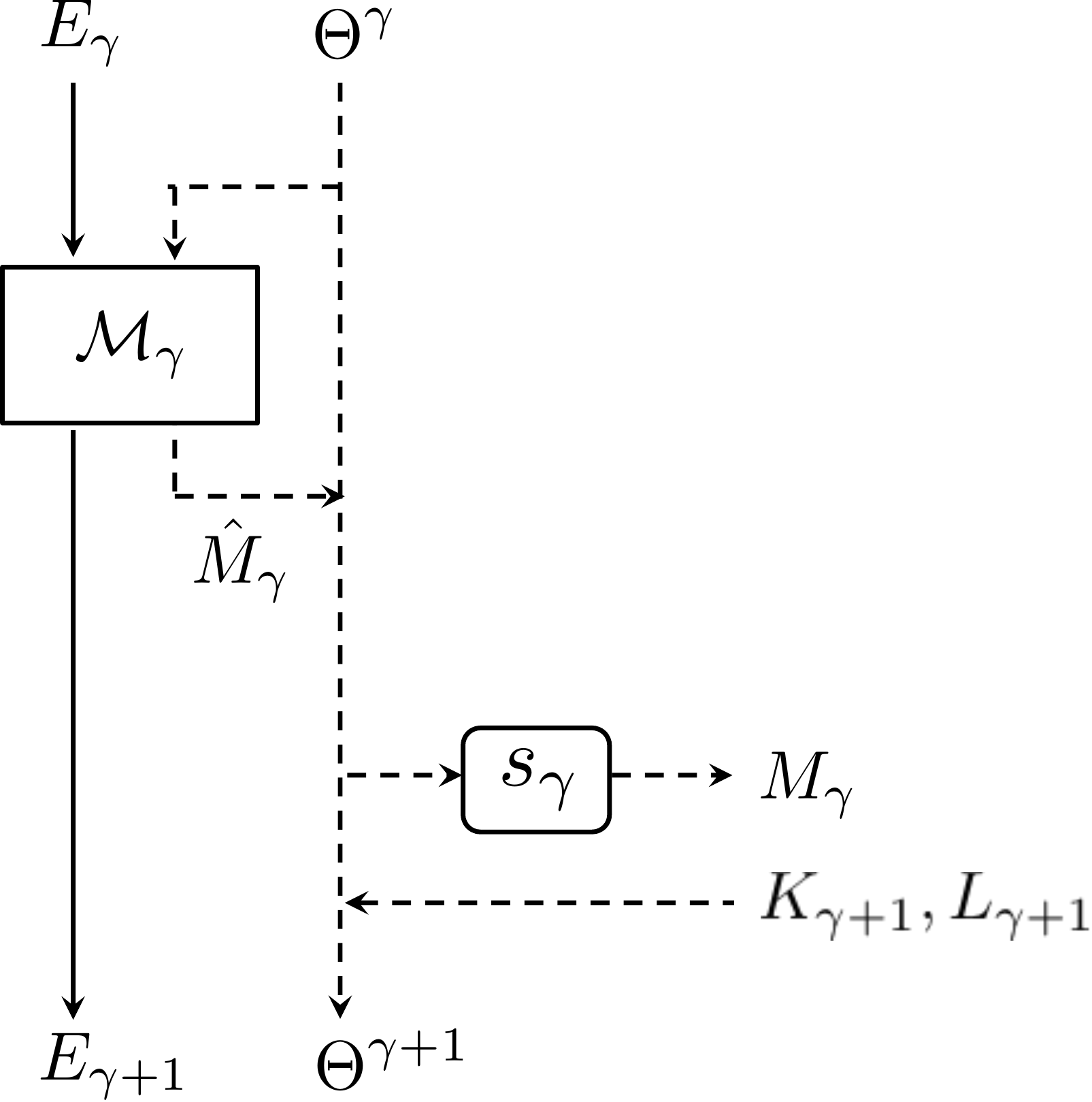}
\end{center}\caption{A graphical representation of the $\gamma$-th step in an LOCC protocol is depicted. We denote system $E$ before the $\gamma$-th step by $E_\gamma$ for $\gamma=1,\cdots,\Gamma$, respectively.}
\label{fig:LOCC}
\end{figure}

We denote by $E_{\gamma}$ and $E_{\gamma+1}$ the input and output systems of $\gamma$-th measurement by Eve, respectively. 
Let us
introduce the following notations:
\begin{eqnarray}
&K^{\gamma}:=(K_1,\cdots,K_{\gamma}),&L^{\gamma}:=(L_1,\cdots,L_{\gamma})\nonumber\\
&M^{\gamma}:=(M_1,\cdots,M_{\gamma}),&{\hat M}^{\gamma}:=({\hat M}_1,\cdots,{\hat M}_{\gamma})\nonumber
\end{eqnarray}
In general, Eve's measurement in the protocol, as well as classical messages that she broadcasts, may dependent on the previous measurement outcomes and messages in the following way (Figure \ref{fig:LOCC}):
\begin{itemize}
\item ${\mathcal M}_\gamma$ depends on $(K^{\gamma},L^{\gamma},{\hat M}^{\gamma-1})$,
\item $M_{\gamma}$ depends on $(K^{\gamma},L^{\gamma},{\hat M}^{\gamma})$.
\end{itemize}
Here, we have defined ${\hat M}^0=\emptyset$. In the following, we denote by $\Theta^{\gamma}$ the triplet of random variables $(K^{\gamma},L^{\gamma},{\hat M}^{\gamma-1})$ and the system in which the variables are registered.

\subsection{Description of Eve's Measurement}

We prove that, without loss of generality, we may assume that the measurement by Eve in each step is a ``noiseless'' measurement. 
To be more precise,
let $\ca{T}_n$ be an $(n,\epsilon)$-protocol for assisted entanglement dilution of $\omega^{AB}$. There exists an $(n,\epsilon)$-protocol $\ca{T}_n'$, which induces the same map as $\ca{T}_n$, such that $C^\downarrow(\ca{T}_n)=C^\downarrow(\ca{T}_n')$ and that the $\gamma$-th measurement by Eve is represented by a CPTP map $\ca{M}_\gamma'$ from $E_\gamma$ to $E_{\gamma+1}M_\gamma$, defined by
\alg{
\ca{M}_\gamma'(\cdot)=\sum_{\hat{m}\in{\mfk M}_\gamma}M_m^\gamma(\cdot)M_m^{\gamma\dagger}\otm\proj{m}^{M_\gamma},
\laeq{cureel}
}
where $\{M_m^\gamma\}_{m\in{\mfk M}_\gamma}$ is the set of measurement operators.

The proof is as follows.
In general, the $\gamma$-th measurement by Eve for a given value $\theta^{\gamma}$ of $\Theta^{\gamma}$ is represented by a CPTP map
\alg{
\ca{M}_\gamma^{\theta^{\gamma}}(\cdot)=\sum_{\hat{m}\in\hat{\mfk M}_\gamma}\ca{M}_{\gamma,\hat{m}}^{\theta^{\gamma}}(\cdot)\otm\proj{\hat{m}}^{\hat{M}_\gamma},
}
where $\ca{M}_{\gamma,\hat{m}}^{\theta^{\gamma}}$ are completely positive maps from $E_{\gamma}$ to $E_{\gamma+1}$ such that $\sum_{\hat{m}=1}^M\ca{M}_{\gamma,\hat{m}}^{\theta^\gamma}(\cdot)$ is trace-preserving. 
The message $M_\gamma$ is obtained by applying a stochastic map on $\Theta^\gamma$ and the measurement outcome $\hat{M}_\gamma$. Thus, by incorporating the message, Eve's operation is represented by a CPTP map
\alg{
\bar{\ca{M}}_\gamma^{\theta^{\gamma}}(\cdot)
=
&
\sum_{\hat{m}\in\hat{\mfk M}_\gamma}
\sum_{m\in{\mfk M}_\gamma}
s_\gamma(m|\hat{m},\theta^{\gamma})
\cdot
\ca{M}_{\gamma,\hat{m}}^{\theta^{\gamma}}(\cdot)
\nn\\
&
\quad\quad\quad\quad
\otm\proj{\hat{m}}^{\hat{M}_\gamma}
\otm\proj{m}^{M_\gamma}.
}
Here, $\{s_\gamma(m|\hat{m},\theta^{\gamma})\}_{m=1}^{|{\mfk M}_\gamma|}$ is a probability distribution that represents the post-processing of the measurement outcome to obtain the message.

Define a map 
$
\bar{\ca{M}}_\gamma:
\Theta^{\gamma}E_{\gamma}
\rightarrow
\Theta^{\gamma}\hat{M}_\gamma M_\gamma E_{\gamma+1}
$
by
\alg{
\bar{\ca{M}}_\gamma(\cdot)
:=
\sum_{\theta^{\gamma}}
\proj{\theta^{\gamma}}^{\Theta^\gamma}
\otm
\bar{\ca{M}}_\gamma^{\theta^{\gamma}}(\bra{\theta^{\gamma}}(\cdot)\ket{\theta^{\gamma}}),
\nn
}
and let $\ca{V}_\gamma:\Theta^{\gamma}E_{\gamma}\rightarrow\Theta^{\gamma}\hat{M}_\gamma M_\gamma E_{\gamma+1}\tilde{E}_\gamma$ be a linear isometry such that a Stinespring dilation of $\bar{\ca{M}}_\gamma$ is given by $\bar{\ca{M}}_\gamma={\rm Tr}_{\tilde{E}_\gamma}\circ\ca{V}_\gamma$. 
Consider a protocol $\ca{T}_n'$ in which local operations and public communication by Alice and Bob are the same as those in $\ca{T}_n$, but the $\gamma$-th measurement by Eve is given by
\alg{
\ca{M}_\gamma'(\cdot):=
\sum_{m\in{\mfk M}_\gamma}
\bra{m}^{M_\gamma}\ca{V}_\gamma(\cdot)\ket{m}^{M_\gamma}
\otm\proj{m}^{M_\gamma}.
\laeq{FinG}
}
Eve discards all of her systems at the end, including the systems $\tilde{E}_1\cdots\tilde{E}_\Gamma$. It is straightforward to verify that the protocol $\ca{T}_n'$ induces the same map as $\ca{T}_n$, i.e., $\ca{T}_n(\rho)=\tilde{\ca{T}}_n(\rho)$ for any initial state $\rho$. 
\QED

\subsection{Simulation of Eve's Measurement by Quantum Communication}

Each measurement by Eve in the protocol $\ca{T}_n'$, defined by \req{FinG}, and the subsequent broadcasting of classical messages to Alice and Bob, can be converted to a reversible operation by Eve, followed by quantum communication from Eve to Alice and broadcasting of a classical message by Alice. To establish this, we adopt a protocol called {\it coherent communication} \cite{harrow04}. 
For each $\gamma$, let $A^*_{\gamma}$ and $E^*_{\gamma}$ be $\lceil\sqrt{\mu_\gamma}\rceil$-dimensional quantum registers possessed by Alice and Eve, respectively,
and let $\ket{\Phi_\gamma}^{A^*_{\gamma}E^*_\gamma}$ be the maximally entangled state thereon. 
Let $\sigma_m\:(m=1,\cdots,\mu_\gamma)$ be a set of unitaries on $\ca{H}^{E^*_{\gamma}}$ that are orthogonal with respect to the Hilbert-Schmidt inner product, i.e.,
\alg{
{\rm Tr}[\sigma_{m_1}^\dagger\sigma_{m_2}]\propto\delta_{m_1,m_2}.
\laeq{orthoPauli}
} 
An example of such a set of unitaries is that of the generalized Pauli operators (see e.g. \cite{bartlett2002quantum}).
Based on $\ca{M}_\gamma'$ defined by \req{cureel}, we introduce the following operation  $\tilde{\ca{M}_\gamma}$ that Eve performs in the $\gamma$-th step:
\benum

\item
After receiving classical messages $K_\gamma$ and $L_\gamma$ from Alice and Bob, Eve performs an isometry $\ca{V}_\gamma$ that satisfies \req{FinG}.

\item
Eve performs a controlled-unitary operation in the form of
\alg{
U_\gamma^{E^*_{\gamma}M_\gamma}:=
\sum_{m=1}^{\mu_\gamma}
\sigma_m^{E^*_{\gamma}}
\otm\proj{m}^{M_\gamma}.
}
\item
Eve transmits the system $E^*_{\gamma}$ to Alice.
\ennum
Note that $\sigma_m\ket{\Phi_\gamma}$ is orthogonal for $m\neq m'$ due to \req{orthoPauli}.
Thus, Alice can perfectly obtain $m$ by performing a measurement on $A^*_{\gamma}E^*_{\gamma}$. The quantum communication cost of this protocol is given by $\log{\dim{E^*_{\gamma}}}=\log{\lceil\sqrt{\mu_\gamma}\rceil}$.

Using $\tilde{\ca{M}_\gamma}$, we construct a protocol $\tilde{\ca{T}}_n\in\Omega_Q$ from $\ca{T}_n$ as follows:
\benum

\item
In $\gamma$-th step, instead of performing a measurement $\ca{M}_\gamma$ and broadcasting a classical message, Eve performs $\tilde{\ca{M}_\gamma}$ and sends $E^*_{\gamma}$ to Alice.
  
\item
In $(\gamma+1)$-th step, before performing the $(\gamma+1)$-th operation in $\tilde{\ca{T}}_n$, Alice performs a measurement on $A^*_{\gamma}E^*_{\gamma}$ to obtain $m$ and broadcasts it to Bob and Eve.
\ennum
It is straightforward to verify that $\ca{T}_n$ and $\tilde{\ca{T}}_n$ induces the same map.

\subsection{Construction of $\tilde{\ca{T}}_{nl}$}

Let $\tilde{\ca{T}}_{nl}\in\Omega_Q$ be an operation that is constructed from $\ca{T}_{nl}:=\ca{T}_n^{\otm l}$, along the same line as we have constructed $\tilde{\ca{T}}_{n}$ from $\ca{T}_n$.
Without loss of generality, we may assume that 
\begin{align}
C_{(\gamma)}^{\rm E \rightarrow A}(\ca{T}_n)+C_{(\gamma)}^{\rm E \rightarrow B}(\ca{T}_n)
\geq
1\nn
\end{align}
for each $\gamma$, in which case we have
\alg{
\Gamma(\ca{T}_n)
\leq
C^\downarrow(\ca{T}_n).
\laeq{oomomo}
}
The quantum communication cost from Eve to Alice in the $\gamma$-th step of $\tilde{\ca{T}}_{nl}$ is given by
\alg{
&
\log{\left\lceil\sqrt{\mu_\gamma^l}\right\rceil}
\leq
\log{\left(\sqrt{\mu_\gamma^l}+1\right)}
\nn\\
&
=
\log{\sqrt{\mu_\gamma^l}}
+
\log{\left(1+1/\sqrt{\mu_\gamma^l}\right)}
\nn\\
&\leq
\frac{l}{2}
\log{\mu_\gamma}+1.
}
Hence, the total quantum communication cost is calculated to be
\alg{
{\rm Qc}(\tilde{\ca{T}}_{nl})
\leq
\sum_{\gamma=1}^{\Gamma(\ca{T}_n)}\left(\frac{l}{2}\log{\mu_\gamma}+1\right)
\leq
\left(\frac{l}{2}+1\right)C^\downarrow(\ca{T}_n),\nn
}
where we have used \req{oomomo} in the second inequality.
This completes the proof of \rLmm{wawawa}.
\QED

\bibliographystyle{IEEEtran}
\bibliography{bibbib.bib}

\begin{thebibliography}{10}
\providecommand{\url}[1]{#1}
\csname url@samestyle\endcsname
\providecommand{\newblock}{\relax}
\providecommand{\bibinfo}[2]{#2}
\providecommand{\BIBentrySTDinterwordspacing}{\spaceskip=0pt\relax}
\providecommand{\BIBentryALTinterwordstretchfactor}{4}
\providecommand{\BIBentryALTinterwordspacing}{\spaceskip=\fontdimen2\font plus
\BIBentryALTinterwordstretchfactor\fontdimen3\font minus
  \fontdimen4\font\relax}
\providecommand{\BIBforeignlanguage}[2]{{%
\expandafter\ifx\csname l@#1\endcsname\relax
\typeout{** WARNING: IEEEtran.bst: No hyphenation pattern has been}%
\typeout{** loaded for the language `#1'. Using the pattern for}%
\typeout{** the default language instead.}%
\else
\language=\csname l@#1\endcsname
\fi
#2}}
\providecommand{\BIBdecl}{\relax}
\BIBdecl

\bibitem{yard2009optimal}
J.~T. Yard and I.~Devetak, ``Optimal quantum source coding with quantum side
  information at the encoder and decoder,'' \emph{IEEE Trans. Inf. Theory},
  vol.~55, no.~11, pp. 5339--5351, 2009.

\bibitem{devetak2008exact}
I.~Devetak and J.~Yard, ``Exact cost of redistributing multipartite quantum
  states,'' \emph{Phys. Rev. Lett.}, vol. 100, no.~23, p. 230501, 2008.

\bibitem{berta2018conditional}
M.~Berta, F.~G. Brand{\~a}o, C.~Majenz, and M.~M. Wilde, ``Conditional
  decoupling of quantum information,'' \emph{Phys. Rev. Lett.}, vol. 121,
  no.~4, p. 040504, 2018.

\bibitem{berta2018deconstruction}
M.~Berta, F.~G. Brandao, C.~Majenz, and M.~M. Wilde, ``Deconstruction and
  conditional erasure of quantum correlations,'' \emph{Phys. Rev. A}, vol.~98,
  no.~4, p. 042320, 2018.

\bibitem{fawzi2015quantum}
O.~Fawzi and R.~Renner, ``Quantum conditional mutual information and
  approximate {Markov} chains,'' \emph{Comm. Math. Phys.}, vol. 340, no.~2, pp.
  575--611, 2015.

\bibitem{hayden04}
P.~Hayden, R.~Jozsa, D.~Petz, and A.~Winter, ``Structure of states which
  satisfy strong subadditivity of quantum entropy with equality,'' \emph{Comm.
  Math. Phys.}, vol. 246, pp. 359--374, 2004.

\bibitem{sharma2017conditional}
K.~Sharma, E.~Wakakuwa, and M.~M. Wilde, ``Conditional quantum one-time pad,''
  \emph{Phys. Rev. Lett.}, vol. 124, no.~5, p. 050503, 2020.

\bibitem{christandl04}
M.~Christandl and A.~Winter, ````squashed entanglement'': An additive
  entanglement measure,'' \emph{J. Math. Phys.}, vol. 45.3, pp. 829--840, 2004.

\bibitem{brandao11}
F.~G. S.~L. Brand{\~a}o, M.~Christandl, and J.~Yard, ``Faithful squashed
  entanglement,'' \emph{Comm. Math. Phys.}, vol. 306, pp. 805--830, 2011.

\bibitem{gisin2000linking}
N.~Gisin and S.~Wolf, ``Linking classical and quantum key agreement: is there
  ``bound information''?'' in \emph{Ann. Int. Crypt. Conf.}\hskip 1em plus
  0.5em minus 0.4em\relax Springer, 2000, pp. 482--500.

\bibitem{christandl2004intrinsic}
M.~Christandl and R.~Renner, ``On intrinsic information,'' in \emph{ISIT 2004.
  Proceedings.}\hskip 1em plus 0.5em minus 0.4em\relax IEEE, 2004, p. 135.

\bibitem{christandl2007unifying}
M.~Christandl, A.~Ekert, M.~Horodecki, P.~Horodecki, J.~Oppenheim, and
  R.~Renner, ``Unifying classical and quantum key distillation,'' in
  \emph{Theory of Crypt. Conf.}\hskip 1em plus 0.5em minus 0.4em\relax
  Springer, 2007, pp. 456--478.

\bibitem{sutter2018approximate}
D.~Sutter, ``Approximate quantum {Markov} chains,'' in \emph{Approximate
  Quantum Markov Chains}.\hskip 1em plus 0.5em minus 0.4em\relax Springer,
  2018, pp. 75--100.

\bibitem{plenio07}
M.~B. Plenio and S.~Virmani, ``An introduction to entanglement measures,''
  \emph{Quant. Inf. Comput.}, vol.~7, pp. 1--51, 2007.

\bibitem{horodecki2009quantum}
R.~Horodecki, P.~Horodecki, M.~Horodecki, and K.~Horodecki, ``Quantum
  entanglement,'' \emph{Rev. Mod. Phys.}, vol.~81, no.~2, p. 865, 2009.

\bibitem{maurer1996towards}
U.~Maurer and S.~Wolf, ``Towards characterizing when information-theoretic
  secret key agreement is possible,'' in \emph{Int. Conf. Theo. App. Crypt.
  Inf. Sec.}\hskip 1em plus 0.5em minus 0.4em\relax Springer, 1996, pp.
  196--209.

\bibitem{maurer1997intrinsic}
------, ``The intrinsic conditional mutual information and perfect secrecy,''
  in \emph{Proc. of 1997 IEEE Int. Symp. Info. Theory}.\hskip 1em plus 0.5em
  minus 0.4em\relax IEEE, 1997, p.~88.

\bibitem{maurer1999unconditionally}
U.~M. Maurer and S.~Wolf, ``Unconditionally secure key agreement and the
  intrinsic conditional information,'' \emph{IEEE Trans. Inf. Theory}, vol.~45,
  no.~2, pp. 499--514, 1999.

\bibitem{renner2003new}
R.~Renner and S.~Wolf, ``New bounds in secret-key agreement: The gap between
  formation and secrecy extraction,'' in \emph{EUROCRYPT}, vol. 2656.\hskip 1em
  plus 0.5em minus 0.4em\relax Springer, 2003, pp. 562--577.

\bibitem{horodecki2005information}
K.~Horodecki, M.~Horodecki, P.~Horodecki, and J.~Oppenheim, ``Information
  theories with adversaries, intrinsic information, and entanglement,''
  \emph{Found. of Phys.}, vol.~35, no.~12, pp. 2027--2040, 2005.

\bibitem{banerjee2015secret}
P.~K. Banerjee, ``A secret common information duality for tripartite noisy
  correlations,'' in \emph{Int. Symp. on Sec. Comp. Comm.}\hskip 1em plus 0.5em
  minus 0.4em\relax Springer, 2015, pp. 329--341.

\bibitem{chitambar2015classical}
E.~Chitambar, B.~Fortescue, and M.-H. Hsieh, ``Classical analog to entanglement
  reversibility,'' \emph{Phys. Rev. Lett.}, vol. 115, no.~9, p. 090501, 2015.

\bibitem{chitambar2015distributions}
------, ``Distributions attaining secret key at a rate of the conditional
  mutual information,'' in \emph{Ann. Crypt. Conf.}\hskip 1em plus 0.5em minus
  0.4em\relax Springer, 2015, pp. 443--462.

\bibitem{chitambar2018quantum}
E.~Chitambar and G.~Gour, ``Quantum resource theories,'' \emph{Rev. Mod.
  Phys.}, vol.~91, no.~2, p. 025001, 2019.

\bibitem{wakakuwa2017operational}
E.~Wakakuwa, ``Operational resource theory of non-{Markovianity},'' \emph{arXiv
  preprint arXiv:1709.07248}, 2017.

\bibitem{tucci2002entanglement}
R.~R. Tucci, ``Entanglement of distillation and conditional mutual
  information,'' \emph{arXiv preprint quant-ph/0202144}, 2002.

\bibitem{nagel2003another}
O.~A. Nagel and G.~A. Raggio, ``Another state entanglement measure,''
  \emph{arXiv preprint quant-ph/0306024}, 2003.

\bibitem{yang2009squashed}
D.~Yang, K.~Horodecki, M.~Horodecki, P.~Horodecki, J.~Oppenheim, and W.~Song,
  ``Squashed entanglement for multipartite states and entanglement measures
  based on the mixed convex roof,'' \emph{IEEE Trans. Inf. Theory}, vol.~55,
  no.~7, pp. 3375--3387, 2009.

\bibitem{wildetext}
M.~Wilde, \emph{Quantum Information Theory}, 2nd~ed.\hskip 1em plus 0.5em minus
  0.4em\relax Cambridge University Press, 2017.

\bibitem{lieb1973proof}
E.~H. Lieb and M.~B. Ruskai, ``Proof of the strong subadditivity of
  quantum-mechanical entropy,'' \emph{J. Math. Phys.}, vol.~14, no.~12, pp.
  1938--1941, 1973.

\bibitem{yang2008additive}
D.~Yang, M.~Horodecki, and Z.~Wang, ``An additive and operational entanglement
  measure: conditional entanglement of mutual information,'' \emph{Phys. Rev.
  Lett.}, vol. 101, no.~14, p. 140501, 2008.

\bibitem{abey09}
A.~Abeyesinghe, I.~Devetak, P.~Hayden, and A.~Winter, ``The mother of all
  protocols: Reconstructing quantum information's family tree,'' \emph{Proc. R.
  Soc. A}, vol. 465, p. 2537, 2009.

\bibitem{ye2008quantum}
M.-Y. Ye, Y.-K. Bai, and Z.~D. Wang, ``Quantum state redistribution based on a
  generalized decoupling,'' \emph{Phys. Rev. A}, vol.~78, no.~3, p. 030302,
  2008.

\bibitem{cover05}
T.~M. Cover and J.~A. Thomas, \emph{Elements of Information Theory (2nd
  ed.)}.\hskip 1em plus 0.5em minus 0.4em\relax Wiley-Interscience, 2005.

\bibitem{bennett96}
C.~H. Bennett, D.~P. DiVincenzo, J.~A. Smolin, and W.~K. Wootters,
  ``Mixed-state entanglement and quantum error correction,'' \emph{Phys. Rev.
  A}, vol.~54, p. 3824, 1996.

\bibitem{anand2019quantifying}
N.~Anand and T.~A. Brun, ``Quantifying non-{M}arkovianity: a quantum
  resource-theoretic approach,'' \emph{arXiv:1903.03880}, 2019.

\bibitem{bhattacharya2018convex}
S.~Bhattacharya, B.~Bhattacharya, and A.~Majumdar, ``Convex resource theory of
  non-{M}arkovianity,'' \emph{arXiv:1803.06881}, 2018.

\bibitem{winczewski2019upper}
M.~Winczewski, T.~Das, and K.~Horodecki, ``Upper bounds on secure key against
  non-signaling adversary via non-signaling squashed secrecy monotones,''
  \emph{arXiv preprint arXiv:1903.12154}, 2019.

\bibitem{kaur2020fundamental}
E.~Kaur, M.~M. Wilde, and A.~Winter, ``Fundamental limits on key rates in
  device-independent quantum key distribution,'' \emph{New J. of Phys.},
  vol.~22, no.~2, p. 023039, 2020.

\bibitem{winter2016tight}
A.~Winter, ``Tight uniform continuity bounds for quantum entropies: conditional
  entropy, relative entropy distance and energy constraints,'' \emph{Comm.
  Math. Phys.}, vol. 347, no.~1, pp. 291--313, 2016.

\bibitem{alicki04}
R.~Alicki and M.~Fannes, ``Continuity of quantum conditional information,''
  \emph{J. Phys. A: Math. Gen.}, vol. 37.5, pp. L55--L57, 2004.

\bibitem{harrow04}
A.~Harrow, ``Coherent communication of classical messages,'' \emph{Phys. Rev.
  Lett.}, vol.~92, p. 097902, 2004.

\bibitem{bartlett2002quantum}
S.~D. Bartlett, H.~de~Guise, and B.~C. Sanders, ``Quantum encodings in spin
  systems and harmonic oscillators,'' \emph{Phys. Rev. A}, vol.~65, no.~5, p.
  052316, 2002.

\end{thebibliography}

\end{document}